\documentclass[twocolumn]{aastex6}
\usepackage{xcolor}
\usepackage{float}
\usepackage[caption = false]{subfig}

\newcommand\ammonia{\mbox{NH}_{3}}
\newcommand\Ek{E_{\mathrm{K}}}
\newcommand\Eg{E_{\mathrm{g}}}
\newcommand\Etot{E_{\mathrm{tot}}}
\newcommand\kms{\mbox{\,km s}^{-1}}
\newcommand\Lcal{{\cal{L}}}

\newcommand\Mstar{M_*}
\newcommand\Msun{M_{\odot}}
\newcommand\pcc{\mbox{\,cm}^{-3}}
\newcommand\sigmaeq{\sigma_{\rm eq}}
\newcommand\sigmag{\sigma_{\rm g}}
\newcommand\sigmanth{\sigma_{\rm nth}}
\newcommand\sigmat{\sigma_{\rm t}}
\newcommand\veldisp{\sigma_{\rm v}}

\newcommand\dendro{\texttt{dendrogram}}
\newcommand\ttmd{{\tt min\_delta}}
\newcommand\ttmv{{\tt min\_value}}
\newcommand\nth{n_{{\rm th}}}

\begin{document} 

\title{Simultaneous evolution of the virial parameter and star formation
rate in molecular clumps undergoing global hierarchical collapse} 

\author{Vianey Camacho \altaffilmark{1}, 
	Enrique V\'azquez-Semadeni\altaffilmark{1}, 
	Aina Palau\altaffilmark{1}, Gemma Busquet\altaffilmark{2,3},
	Manuel Zamora-Avil\'es\altaffilmark{4}}

\altaffiltext{1}{Instituto de Radioastronom\'ia y Astrof\'isica, UNAM, Apartado 
Postal 3-72, 58089 Morelia Michoac\'an, M\'exico}
\altaffiltext{2}{Institut de Ciències de l’Espai (ICE, CSIC), Can Magrans, s/n, 08193 Cerdanyola del Vallès, Catalonia, Spain} 
\altaffiltext{3}{Institut d’Estudis Espacials de Catalunya (IEEC), 08034 Barcelona, Catalonia, Spain}	
\altaffiltext{4}{CONACYT-Instituto Nacional de Astrof\'isica, \'Optica y Electr\'onica, Luis E. Erro 1, 72840 Tonantzintla, Puebla, M\'exico}

\begin{abstract} 

We compare dense clumps and cores in a numerical simulation of 
molecular clouds (MCs) undergoing global hierarchical collapse (GHC) 
to observations in two MCs at different evolutionary stages, the Pipe 
and the G14.225 clouds, to test the ability of the GHC scenario to follow 
the early evolution of the energy budget and star formation activity of 
these structures.
In the simulation, we select a region that contains cores of 
sizes and densities similar to the Pipe cores, and find that it evolves 
through accretion, developing substructure similar to that of
G14.225 cloud after $\sim 1.6$ Myr. 
Within this region, we follow the evolution of the Larson ratio 
$\Lcal \equiv \veldisp/R^{1/2}$, where $\veldisp$ is the velocity dispersion 
and $R$ is the size, the virial parameter $\alpha$, and the star formation 
activity  of the cores/clumps. 
In the simulation, we find that as the region evolves :
{\it i)} its clumps have $\Lcal$ and $\alpha$ values first consistent with 
those of the Pipe substructures and later with those of G14.225;
{\it ii)} the individual cores first exhibit a decrease in $\alpha$ followed 
by an increase when star formation begins;
{\it iii)} collectively, the ensemble of cores/clumps reproduces the observed 
trend of lower $\alpha$ for higher-mass objects, and
{\it iv)} the star formation rate and star formation efficiency increase 
monotonically.
We suggest that this evolution is due to the simultaneous loss of 
externally-driven compressive kinetic energy and increase 
of the self-gravity-driven motions. 
We conclude that the GHC scenario provides a realistic description 
of the evolution of the energy budget of the clouds' substructure at early 
times, which occurs simultaneously with an evolution of the star formation 
activity.

\end{abstract}

\keywords{ISM: clouds -- ISM: kinematics and dynamics -- stars: formation}


\section{Introduction}\label{s:intro} 

One of the most studied parameters of molecular cloud (MC) structure 
is the so-called {\it virial parameter} $\alpha$, defined as the ratio of 
twice the kinetic energy,  $\Ek$, to the gravitational energy, $\Eg$, for 
a uniform-density sphere \citep{Bertoldi92},
\begin{equation}
\alpha \equiv \frac{2\Ek}{|\Eg|} = \frac{5 \veldisp^2 R}{GM}, 
\label{eq:alpha}
\end{equation}
where $\veldisp$ is the average one-dimensional velocity dispersion 
along the line of sight, $R$ is the characteristic radius of the cloud, 
and $M$ its mass.

The standard notion of MCs is that they are quasi-virialized structures, 
in which their self-gravitational energy is globally balanced by the 
turbulent energy \citep[e.g.,] [] {Larson81, MacLow04, McKee07, 
Ballesteros07, Heyer09}\footnote{Magnetic support has recently lost 
appeal because it now appears that MCs tend to be generally 
magnetically supercritical, and thus cannot be globally supported by 
magnetic fields \citep[e.g.,] [] {Crutcher12}}. In this case, the virial 
parameter of MCs in general should be $\sim1$. Observationally, 
however, the virial parameter of MCs and their substructures 
(parsec-scale clumps and 0.1-pc scale cores) appears to be 
significantly larger than unity for clumps of low column density or low 
mass, and to decrease systematically to values smaller than unity for 
objects of higher column density, or higher mass \citep[e.g.,] []{Kauffmann13, 
Leroy15, Liu15, Ohashi16, Sanhueza17, Contreras18, Traficante18a, Louvet18}.

The large values of $\alpha$ (significantly larger than unity) observed 
in clouds of low column densities or masses are often interpreted in 
terms of the presence of a large external confining pressure 
\citep[ P/$k_{\rm B} \sim 10^4 - 10^6$ K$\pcc$; e.g.,][]{Keto86, Oka01,
Field11, Leroy15, Traficante18a}, although it is hard to imagine that such 
high pressures can be thermal in general, since the mean ambient 
thermal pressure in the ISM is rather low, $\sim 3-4 \times 10^3$ K$\pcc$, 
and large deviations from it occur very infrequently \citep[e.g., ][]{Boulares90, 
Jenkins04, Jenkins11}. Instead, it is most likely that these values correspond 
to ram pressure, in which case they imply mass, momentum and energy 
flux across Eulerian cloud boundaries, or a displacement of Lagrangian 
boundaries \citep{Ballesteros99, Banerjee09}. Indeed, in a previous study 
\citep{Camacho16}, we find, through measurement of the mean velocity 
divergence within the clouds in numerical simulations of cloud formation 
and evolution, that roughly half the clouds with an excess of kinetic energy 
are undergoing compression. This can be interpreted as the clouds being 
subject to a ram pressure (which amounts to an inertial compression) that 
is making them denser and smaller, so that they eventually will become 
gravitationally bound.
The origin of this ram pressure can be large scale turbulence, a large 
scale potential well or other instabilities. Furthermore, one important 
possibility is that clouds may be falling into the potential well of a stellar 
spiral arm, which is the main source of large-scale compression for the gas 
in the Galactic disk \citep{Roberts69}.
Thus, this is not really a ``confinement'', since the clouds are not at rest. 
The same goes for the other half of the clouds, which are undergoing 
expansion. In this case, the excess of kinetic energy corresponds to the 
expansion motions, and again the cloud is not confined, so there is no 
need for a high confining pressure. 
In \citet{Camacho16} and \citet[] [hereafter BP+18] {Ballesteros18} it has 
been suggested that, for clouds formed by  inertial compressions in the 
background medium \citep{Ballesteros99}, and which gradually become 
more strongly gravitationally bound, while the inertial compressive motions 
decay or dissipate, the kinetic energy transits from being dominated by the 
inertial motions to being dominated by the gravitationally-driven motions 
\citep[see also][]{Collins12}. 
In that case, an initial decay of the Larson ratio and the virial parameter 
may be expected.

On the other hand, values of $\alpha$ smaller than unity have been 
interpreted as either being in a state of collapse and/or support from 
strong magnetic fields \citep[e.g.,] []{Kauffmann13, Liu15, Ohashi16, 
Sanhueza17, Contreras18}. However, BP+18 recently proposed that 
values of $\alpha < 1$ may be expected in cores that have just recently 
decoupled\footnote{As explained in \citet{VS19}, in the Global Hierarchical 
Collapse (GHC) scenario, clouds are collapsing globally and therefore 
the mean Jeans mass in them decreases with time. As a consequence,  
turbulent density fluctuations (cores) ``decouple" from the general cloud 
flow and begin to collapse themselves when the average Jeans mass in 
the cloud becomes smaller than their own mass.} from the general cloud 
flow and begun to collapse locally if the initial inertial motions are also 
smaller than the virial value. This can be seen by assuming that the 
non-thermal contribution to the velocity dispersion, $\sigmanth$, consists 
itself of two contributions, one being a gravitationally-driven infall velocity
$\sigmag$, and the other a truly turbulent (or inertial; i.e., not consisting of 
infall motions) one dimensional component\footnote{Note that this assumption 
differs form the very common one that the non-thermal component of $\sigma$ 
is due exclusively to the turbulent motions. However, under the assumption 
in GHC that there are infall motions at all scales in the cloud, it becomes 
necessary to distinguish the turbulent and the infall contributions to the 
non-thermal velocity dispersion.}, $\sigmat$, so that 

\begin{equation}
\sigmanth^2 = \sigmag^2 + \sigmat^2.
\label{eq:sigmatot}
\end{equation}

Next, BP+18 pointed out that, when a core of fixed mass  $M$ begins to 
contract locally, it does so from a finite radius $R_0$.  Thus, its 
gravitationally-driven velocity $\sigmag$ at a later, smaller radius $R$, is given by
the condition $\Ek + \Eg = \Etot$, where 

\begin{equation}
\Ek =\frac{1}{2}M\sigmag^2
\end{equation}
and
\begin{equation}
\Eg = - \eta \frac{GM^2} {R},~~~~~~~~ \Etot = -\eta \frac{GM^2} {R_0},
\end{equation}
where $\eta$ is a parameter of order unity that depends on the
geometry of the cloud. Thus, 
\begin{equation}
\sigmag = \sqrt{2\eta GM \left(\frac{1}{R} - \frac{1}{R_0} \right)}.
\label{eq:vinfall}
\end{equation}
As a consequence, the contribution of 
$\sigmag$ to the so-called {\it Larson ratio},
\begin{equation}
 \Lcal \equiv \frac{\veldisp }{R^{1/2}} 
\label{eq:larsonr}
\end{equation}
in this fixed-mass core scales with column density as
\begin{equation}
\Lcal_{\rm g} = \sqrt{ 2\pi \eta G\Sigma \left[1 -
      \left(\frac{\Sigma_0}{\Sigma}\right)^{1/2} \right]},
\label{eq:Lcal-Sigma}
\end{equation}
while its contribution to the virial parameter scales as
\begin{equation}
\alpha_{\rm g} = 2 \left[1 - \left(\frac{\Sigma_0} {\Sigma} \right)^{1/2}
\right],
\label{eq:alpha-Sigma}
\end{equation}
where $\Sigma_0$ is the column density of the core when it began its
contraction. 

On the other hand, simultaneously with the variation of $\sigmag$ with 
column density as the core collapses, BP+18 assumed that the inertial 
contribution to the Larson ratio and the virial parameter decreases by 
dissipation during the compression at two different plausible rates. Thus, 
the total value of $\Lcal$, obtained in quadrature from the gravitational 
and the inertial contributions, may adopt a variety of shapes depending 
on the initial ratio of the two contributions, as shown in Fig.\ 1 of BP+18.
 
Another representation of the total virial parameter can be inferred from 
eqs. (\ref{eq:alpha}) and (\ref{eq:sigmatot}). From this, it follows that 
$\alpha_{\rm tot} = \alpha_{\rm t} + \alpha_{\rm g}$. Thus, according to 
Eq. (\ref{eq:vinfall}), the virial parameter is   
\begin{equation}
\alpha =  \frac{3 \sigma_\mathrm{ t}^2 R} {\eta G M} +
		2\left(1-\frac{R}{R_0} \right).
\end{equation}
In this equation the first term can be recognized as the common definition 
of the virial parameter, while the second term represents the gravitational 
contribution that depends on size.

It is important to note that, in the above treatment, a Lagrangian 
definition of the core was used, so that, by construction, the core 
has a constant mass, and thus this simple calculation cannot predict 
a dependence of the virial parameter with mass. However, if the core 
is instead defined in terms of a density or column-density threshold, 
as is standard for cores defined in terms of molecular-line tracers and 
common in numerical simulations \citep[e.g., ] [] {VS97, Ballesteros02, 
Galvan07, Naranjo15, Camacho16, Ibanez16}, then the mass of a 
gravitationally contracting core increases with time together with its 
column density \citep{Naranjo15}. Thus, we expect the virial parameter 
of populations of cores defined by molecular-line tracers to depend on 
mass as well.

It is important to remark that this kind of evolution occurs in the scenario 
of Global Hierarchical Collapse (GHC), in which cores begin to collapse 
locally within a larger cloud that is itself Jeans unstable as well, so that 
the evolution consists of a multi-scale, hierarchically nested, sequence 
of collapses \citep{VS09, Naranjo15, VS17, VS19}. This is important 
in order to allow accretion from the clump onto the core \citep{Naranjo15}. 
In this scenario, the star formation rate (SFR) of the clouds and their
substructure also increases as the cloud globally contracts, until the 
feedback from massive stars, which appear late (after $\sim 5$ Myr) in the 
evolution of the cloud, begins to destroy it, either by dispersing or 
evaporating the dense gas. At this point, the cloud's SFR begins to 
decrease again \citep{Zamora12, Zamora14, Lee16,VS17, VS18, Caldwell18}.

\citet{Ballesteros18} showed that, in isothermal numerical simulations 
of driven turbulence at the parsec scale, the Larson ratio of cores 
evolves in the $\Lcal$ {\it vs.}  $\Sigma$ diagram. 
In the present paper, we show this evolution, as well as that of the virial 
parameter, in a larger-scale, multi-phase simulation (of size 256 pc) of 
giant molecular cloud formation and its subsequent gravitational contraction, 
and show that it occurs simultaneously with an evolution of the star formation 
activity of the cores. With the aim to test the GHC hypothesis that the kinetic 
energy in the cloud and its substructures is dominated by self-gravity, we do 
not include stellar feedback in the simulation.
We also search for the signature of this simultaneous evolution of the virial 
parameter and the SFR in observational data by comparing the location 
of the simulated cores and those in star-forming regions of presumably different 
evolutionary stages, in $\Lcal$ {\it vs.} $\Sigma$ and $\alpha$ {\it vs.} $M$ 
diagrams, and show that there is good qualitative agreement, thus supporting 
the evolutionary nature of the GHC scenario for molecular clouds.

The plan of the paper is as follows. In Section \ref{s:data} we briefly 
describe the simulation and the data from the literature. Next, 
Section \ref{s:method} presents the procedure to derive the physical 
parameters for both the numerical and the observational data. The results 
about the evolution and the energy budget for both samples are reported 
in Section \ref{s:results}. Finally, we discuss our findings in Section 
\ref{s:discussion} and a brief summary in Section \ref{s:summary}.

\begin{figure*}[ht]
    \centering
           \includegraphics[width=0.8\textwidth]{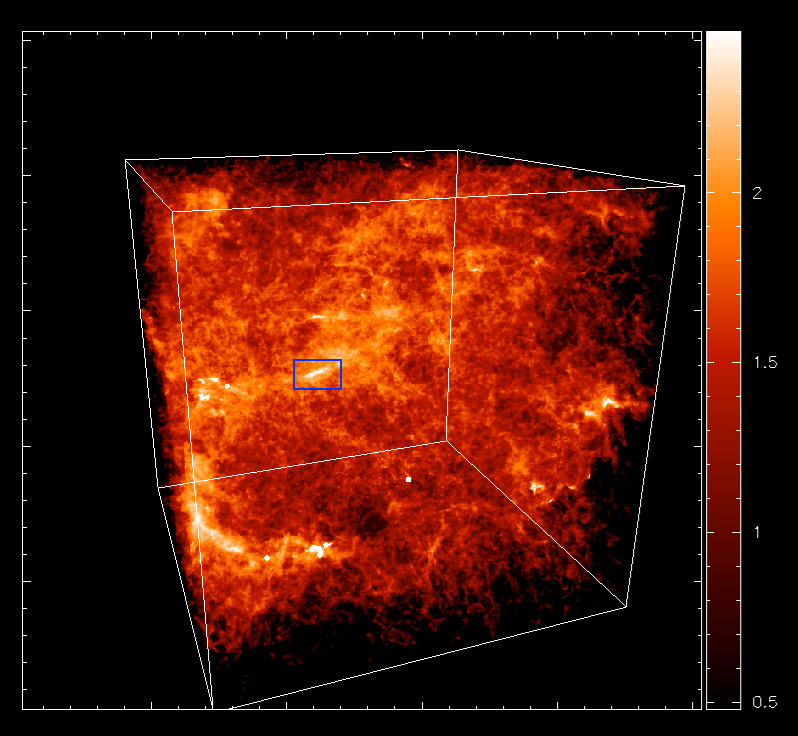} 
    \caption{The simulation after 20.45 Myr of evolution. The box spans 
    256 pc on a side, and contains a total mass $1.5 \sim 10^6 \Msun$. 
    The blue rectangle shows the region studied in this paper, hereafter 
    referred to as Region A. The white dots represent sink particles, 
    which are still scarce at the time shown. The color bar indicates the 
    column density in internal code units.}
  \label{fig:boxsim}
\end{figure*}
%


 \section{Data}\label{s:data} 

\subsection{Observational data}
 
As described in Sec.\ \ref{s:intro}, in order to search for evolutionary 
effects in clouds, we selected two star-forming regions likely to be in 
different evolutionary  stages.  
According to the model of \citet[] [see 
also V\'azquez-Semadeni et al. 2018] {Zamora14}, this evolution 
should manifest itself in different values of the cloud's star formation 
rate and efficiency. Thus, we consider the Pipe Nebula \citep{Alves07,
Rathborne08,Lada08}, a quiescent dark cloud,  and G14.225-0.506 
\citep[][]{Busquet13, Lin17}, an infrared dark cloud (IRDC) showing 
active star formation \citep{Povich16}.

The Pipe Nebula has been studied in both dust extinction 
\citep[e.g.,][] {Lombardi06,Alves07,Roman10,Ascenso13} and molecular 
line emission \citep[C$^{18}$O, $\ammonia$, CCS, HC$_{5}$N; e.g.,][]{
Muench07,Rathborne08,Frau10,Frau12}. It is located at a distance of 
$\sim $145 pc \citep{AlvesF07}, and has a mass $\sim 8000\ \Msun$ 
defined out to an extinction $A_{\rm V}=0.1\ \mathrm{mag}$ \citep{Lada10} 
and a size $\sim 3\times14$ pc \citep{Lada08}. A large population of
dense cores has been identified in this cloud with masses 
$0.2$--$20\Msun$ \citep{Rathborne08,Frau12}.  We selected the sample 
of the cores with $\ammonia$ emission.

IRDC G14.225-0.506 \citep[hereafter G14 for short;] [] {Busquet13}, 
is part of a large molecular cloud  which is actively forming stars 
\citep[][ and references therein]{Povich10,Busquet16, Chen19, 
Shimoikura19}. G14 has a mass $\sim 1.9 \times 10^4 \Msun$
\citep{Lin17}, size $\sim 4.7\times8.7$ pc \citep{Busquet13}, and 
is located at a distance $\sim 1.98$ kpc \citep{Xu11}. In \citet[][
hereafter B+13]{Busquet13}, a study of this cloud was presented 
in the $\ammonia$ (1,1) and (2,2) lines, resulting from a combination 
of Very Large Array (VLA) and Effelsberg 100 m telescope observations. 
We refer the reader to B+13 for details on the observations and data 
reduction. In addition, for consistency with the available data from 
the Pipe Nebula, the FIR/submm-derived column density map towards 
the G14 cloud was also considered \citep{Lin17}.  The dust emission 
map is the result from a combination of ground-based and space 
telescope observations resulting in an angular resolution 
$\sim 10 ''$, which is comparable to the resulting synthesized beam 
of $\ammonia$ in B+13, $\sim 8''\times 7''$.

We assumed that these two clouds represent different stages of 
molecular cloud evolution, due to their different levels of star formation 
activity, as proposed in \citet{VS18}, so that they can be compared to 
different temporal snapshots of the simulation (Sec. \ref{sec:resultobs}).

\subsection{Numerical simulation}

\begin{table*}[]
\centering
\begin{tabular}{cccccccccccc}
\hline
\multicolumn{12}{c}{Pipe sample}                                                                                                                                                                                                                                                                                                                                                                                                                                                \\ \hline
\multicolumn{1}{c}{ID$^{\tt a}$} & \multicolumn{1}{c}{\begin{tabular}[c]{@{}c@{}}R$^{\tt b}$\\ (pc)\end{tabular}} & 
\multicolumn{1}{c}{\begin{tabular}[c]{@{}c@{}}M$^{\tt a,b}$\\ ($\Msun$)\end{tabular}} &
\multicolumn{1}{c}{\begin{tabular}[c]{@{}c@{}}$\sigma_{v,1D}^{\tt a, c}$\\ ($\kms$)\end{tabular}}&
\multicolumn{4}{c}{} &
\multicolumn{1}{c}{ID$^{\tt a}$} & \multicolumn{1}{c}{\begin{tabular}[c]{@{}c@{}}R$^{\tt b}$\\ (pc)\end{tabular}} & 
\multicolumn{1}{c}{\begin{tabular}[c]{@{}c@{}}M$^{\tt a,b}$\\ ($\Msun$)\end{tabular}} &
\multicolumn{1}{c}{\begin{tabular}[c]{@{}c@{}}$\sigma_{v,1D}^{\tt a, c}$\\ ($\kms$)\end{tabular}} \\ \cline{1-4} \cline{9-12}
6	& 	0.12 	&	3.14	&	0.09	&  & & & &    65	 &	0.06  &	0.72	& 	0.26    \\
7	& 	0.14	&	4.69	&	0.08 &  \multicolumn{4}{c}{}  &    70	 & 	0.08	&  	1.14	 & 	0.23    \\
8	& 	0.12	& 	3.26	& 	0.11	&  \multicolumn{4}{c}{}  &    87	 & 	0.17	&      10.3	& 	0.14    \\
12	&	0.23	&	20.3	&	0.15	&  \multicolumn{4}{c}{}  &    89	 & 	0.09	&      1.36	& 	0.10    \\                                                                  
14	& 	0.17	&	9.73	&	0.14 &  \multicolumn{4}{c}{}  &	  91	 & 	0.07	&      1.09	& 	0.07    \\
15	& 	0.12	& 	2.64	& 	0.18 &  \multicolumn{4}{c}{}  &	  92	 & 	0.09	& 	1.61	&	0.19    \\
17	& 	0.07	& 	0.69	& 	0.25 &  \multicolumn{4}{c}{}  &    93	 & 	0.12	& 	3.55	& 	0.17    \\
20	&	0.11	& 	2.28	& 	0.17 &  \multicolumn{4}{c}{}  &    97	 & 	0.18	&	5.86	&	0.21    \\
22	& 	0.08 & 	1.01	&	0.12 &  \multicolumn{4}{c}{}  &	  101	 & 	0.08	& 	1.87	&	0.09    \\
23	& 	0.16	& 	1.87	& 	0.07 &  \multicolumn{4}{c}{}  &	  102	 & 	0.19	& 	6.71	 & 	0.24    \\
25	& 	0.09	& 	1.10	& 	0.20 &  \multicolumn{4}{c}{}  &	  108	 & 	0.08	& 	0.78	& 	0.16    \\
40	& 	0.19	& 	9.23	& 	0.10 &  \multicolumn{4}{c}{}  &	  109	 & 	0.12	& 	3.63	& 	0.08    \\
41	& 	0.08	& 	1.08	& 	0.12 &  \multicolumn{4}{c}{}  &	  113	 & 	0.10	& 	2.39	& 	0.06    \\
42	&	0.09	& 	2.79	& 	0.11 &  \multicolumn{4}{c}{}  &	  132	 & 	0.15	& 	4.67	& 	0.18     \\
47	& 	0.09	& 	1.41	& 	0.14 &  \multicolumn{4}{c}{}  &		&	&	&	\\
\hline
\end{tabular}
\caption{Physical properties for the selected sample of the Pipe cores.  
{\tt a)}\citet{Rathborne08}, 
{\tt b)} \citet{Lada08} 
{\tt c)} $\sigma_{v,1D}$ from the  $\ammonia$(1,1) emission.} 
\label{tab:pipe}
\end{table*}

In order to test the GHC scenario, we choose a numerical simulation 
in which the clouds are born as a consequence of turbulence in the diffuse 
atomic medium, and then engage in global hierarchical collapse as they 
become dominated by self gravity.
This simulation, hereafter RUN03, has been studied in previous works 
\citep{Heiner15, Camacho16} and was performed with the GADGET-2 code 
\citep{Springel05} in a box of 256 pc per side containing 
$296^3 \approx 2.6 \times 10^7$ SPH particles. 
Figure \ref{fig:boxsim} shows the whole  computational domain of the simulation, 
20.45 Myr after the start of the simulation, and just before the first timestep 
considered in our analysis below. The blue rectangle shows the region where 
our numerical cloud is located, illustrating how it fits in the global context of the 
turbulent ISM in the simulation. At this time, star formation is still mild throughout 
the simulation, and has not started yet in the region under consideration. 

The simulation was initialized with a turbulent driver during the first 0.65 Myr 
applied at scales from 1 to 1/4 of the numerical box size, reaching a peak 
velocity dispersion $\sim18 \thinspace \kms$ at that time, after which 
the simulation was left to decay. It also includes a prescription for the 
formation of sink particles \citep[see][for details]{Heiner15}, 
and the fix by \citet{Abel11} that eliminates several unphysical effects of 
the SPH scheme.
These initial conditions result in a clumpy medium 
that evolves self-consistently from long before the clouds become massive 
enough to be considered molecular. During this evolution, the clouds grow in mass, 
size and density by accretion from the diffuse environment, driven first by the inertial 
flow from the turbulence, and later by their self-gravity. The clouds in the simulation 
begin to globally contract gravitationally at $t \sim 10$ Myr after the start of the 
simulation, and sink particles start forming at $t \sim 16$ Myr.
This simulation was evolved for a total of $\approx 34$ Myr.
%


 \section{Methodology}
\label{s:method} 

In order to meaningfully compare the numerical and the observational 
data, we need to carefully select the region to be studied in the 
simulation and, for the observations, to ensure consistency between 
the two datasets. In the simulation, we choose to study a single 
star-forming region at various times in order to determine its energy 
budget evolution. The region was studied using various density thresholds 
to define its internal structure, in order to explore objects from the scale 
of molecular clouds to that of dense cores. For the observational data, 
in the case of the Pipe, we consider a core sample previously identified in 
the literature (see description below), while for G14 we create our own 
sample of filaments, clumps and cores directly from the maps.

\subsection{The Pipe sample } 
\label{sec:pipesample}

The Pipe Nebula has more than a hundred identified dense cores 
\citep{Lada08}. Because line emission measurements provide the 
kinematic information, in this work we selected those cores that 
have been detected  in $\ammonia$ \citep{Rathborne08}, from 
which the velocity dispersion is obtained. The mass of these cores 
is reported in \citet{Lada08}, and spans a range of $\sim 0.5 -20\ \Msun$. 
These masses have been determined from the extinction maps, as 
is the size, which has been computed assuming spherical geometry 
given the area in the plane of the sky, so that $R = (A/\pi)^{1/2}$ \citep{Lada08}. 
The range in size for the selected cores is $\sim 0.06-0.3\ $pc and 
the mean density of the cores is $\sim 7\times10^3 \pcc$ \citep[see 
Table 2 of][]{Rathborne08}. We use the data reported in these works 
for our analysis. Table \ref{tab:pipe} shows the data for the selected 
sample and Fig.\ \ref{fig:obsporperties} shows the correspondent mass-size relation. 

\begin{table*}[]
\centering
\begin{tabular}{cccccccccccc}
\hline
\multicolumn{12}{c}{G14 sample}                                                                                                                                                                                                                                                                                                                                                                                                                                                \\ \hline
\multicolumn{1}{c}{ID$^{\tt a}$} & \multicolumn{1}{c}{\begin{tabular}[c]{@{}c@{}}R$^{\tt b}$\\ (pc)\end{tabular}} & 
\multicolumn{1}{c}{\begin{tabular}[c]{@{}c@{}}M$^{\tt c}$\\ ($\Msun$)\end{tabular}} &
\multicolumn{1}{c}{\begin{tabular}[c]{@{}c@{}}$\sigma_{v,1D}^{\tt d}$\\ ($\kms$)\end{tabular}}&
\multicolumn{4}{c}{ } &
\multicolumn{1}{c}{ID$^{\tt a}$} & \multicolumn{1}{c}{\begin{tabular}[c]{@{}c@{}}R$^{\tt b}$\\ (pc)\end{tabular}} & 
\multicolumn{1}{c}{\begin{tabular}[c]{@{}c@{}}M$^{\tt c}$\\ ($\Msun$)\end{tabular}} &
\multicolumn{1}{c}{\begin{tabular}[c]{@{}c@{}}$\sigma_{v,1D}^{\tt d}$\\ ($\kms$)\end{tabular}} \\ \cline{1-4} \cline{9-12}
C1    &  0.079 & 	8.519 & 	0.437 &  & & & & C11	&  0.191 & 	214.234 & 	0.955 \\
C2	&  0.084 & 	7.828 & 	0.471 &  \multicolumn{4}{c}{}  & C12	&  0.124 & 	47.156 & 	0.556 \\
C3	&  0.109 & 	22.760 & 	0.522 &  \multicolumn{4}{c}{}  & C13	&  0.213 &  	204.163 & 	0.972 \\
C4	&  0.107 & 	20.686 & 	0.803 &  \multicolumn{4}{c}{}  & c$_{\rm o}$1	&  0.092 & 	80.153 & 	0.892 \\
C5	&  0.054 & 	6.169 & 	0.773 &  \multicolumn{4}{c}{}  & c$_{\rm o}$2	&  0.099 & 	66.107 & 	0.981 \\
C6	&  0.056 & 	6.494 & 	0.573 &  \multicolumn{4}{c}{}  & f1	&  0.511 & 	636.414 & 	1.096 \\
C7	&  0.096 & 	21.406 & 	0.544 &  \multicolumn{4}{c}{}  & f2	&  0.488 & 	648.723 &   0.811 \\
C8	&  0.057 & 	6.741 & 	0.437 &  \multicolumn{4}{c}{}  & f3	&  0.274 & 	98.919 & 	0.582 \\
C9	&  0.056 & 	8.163 & 	0.561 &  \multicolumn{4}{c}{}  & f4	&  0.509 & 	574.053 & 	1.333 \\
C10	&  0.060 & 	8.784 & 	0.790 &  \multicolumn{4}{c}{}  &		&	&	&	\\
\hline
\end{tabular}
\caption{Physical properties for the G14 sample. 
 {\tt a)} The IDs in this table correspond to the labeled objects in Fig. (\ref{fig:g14sample}). 
 {\tt b)} The size was computed from the area $A$ defined with the \dendro\ package, $R=(A/\pi)^{1/2}$.
 {\tt c)} Mass is computed  as $M=2.8 \ m_{\rm H}\ N({\rm H}_2)\ A$, considering the $N({\rm H}_2)$ map from \citet{Lin17} and the hydrogen mass $m_{\rm H} =1.6\times10^{-24}$ g.
 {\tt d)} $\sigma_{v,1D}$ was obtained from the  $\ammonia$(1,1) hyperfine fits.
 } 
\label{tab:g14}
\end{table*}

\subsection{The G14 sample} 
\label{subsec:g14procedure}

In the case of G14 data, we created a clump ensemble by identifying 
the structures with the \dendro\footnote{This research made use of Astrodendro, 
a Python package to compute dendrograms of Astronomical data 
(\url{https://github.com/dendrograms/astrodendro})} 
algorithm \citep{Rosolowsky08, Goodman09} applied to the moment-0 map of the $\ammonia$ data.  
In B+13, two classes of objects, filaments and hubs, were recognized. 
The former were identified in the $\ammonia (1,1)$ map, while the latter 
were identified in the $\ammonia (2,2)$ map, and correspond to the sites 
where filaments converge. This indicates that the hubs have larger typical 
densities than the filaments. It is noteworthy however, that the structures 
they call ``hubs" actually still exhibit elongated morphologies.

We make use of the \dendro\ algorithm in the ammonia data to define the boundaries of the 
dense structures similar to those reported in B+13. However, we defined 
three classes of objects: filaments, clumps and dense cores\footnote{We 
refrain from using the name ``hub"  for the structures in the simulation in 
order to avoid confusion.}. Our filaments roughly coincide with those 
defined in B+13; our clumps include two classes of objects: the hubs from 
B+13 as well as some isolated roundish clumps away from the filaments. 
Finally, our dense cores correspond to the densest, roundish regions within 
B+13's hubs.
According to this classification, we chose different sets 
of values of the input parameters in the \dendro\ algorithm, applied to 
the moment-0 map, to obtain the set of pixels defining each structure. 
The rms noise of the $\ammonia$ cube in G14 is 
8 mJy/beam per 0.6 $\kms$ spectral chanel. B+13 defined the lowest 
contour level at 3 times the rms noise; thus, we set the noise parameter 
in the dendrogram algorithm to $\sigma = 3 \thinspace \mathrm{rms}$. 
The  equivalent area of the beam is $\sim 12$ pixels ($\sim 0.1$ pc). 
Then, we consider a minimum number of pixels of \texttt{min\_pix} = 15, 
in order to have structures larger or equal to the beam. Finally, we vary 
the input parameters by means of \ttmv\ = \texttt{n}$_{\rm v} \sigma$ 
and \ttmd = \texttt{n}$_{\rm d} \sigma$. Where 
(\texttt{n}$_{\rm v} $, \texttt{n}$_{\rm d}$) = (1, 1) for the filaments, (5, 1) 
for the clumps, and (15, 3)  for the dense cores. 

\begin{figure}[ht!]
\includegraphics[trim={2cm 9cm 2cm 1.5cm},clip, scale=0.55]{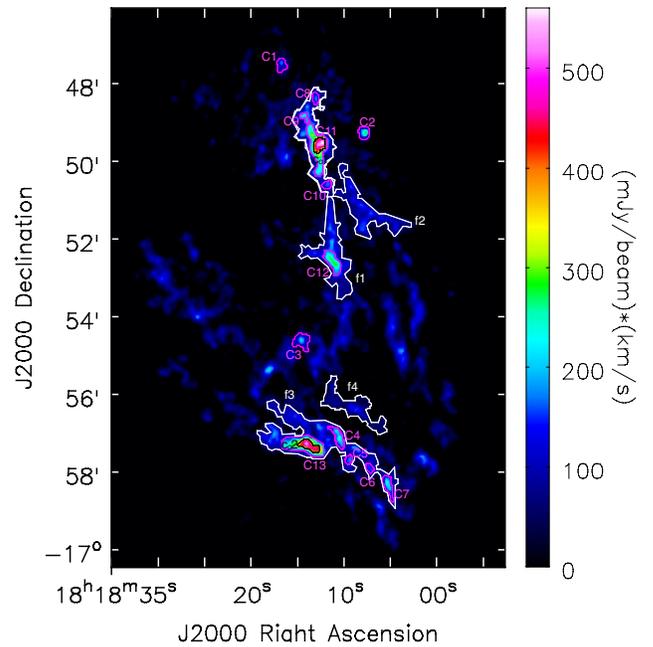}
\caption{ G14 moment-0 map of the $\ammonia$(1,1) inversion transition 
taken from \citet{Busquet13}. Contours show the objects of study in this
work, the different colors denote the filaments-f (white), clumps-C 
(magenta) and dense cores-c$_{\rm o}$ (black). They have been obtained 
using the \dendro\ package.}
\label{fig:g14sample}
\end{figure}

Figure \ref{fig:g14sample} shows the three main families that have been 
selected, the filaments (white) labeled with an ``f", the clumps (magenta) 
labeled with a ``C", and the dense cores (black) labeled with a ``c$_{\rm o}$".
In order to investigate how the physical properties of the sample change when  the \dendro\ input parameters are modified, Appendix \ref{app:dendro} shows a new sample obtained varying  the \ttmd\ value.

With the area defined by the set of pixels for each structure, we computed 
their size.
We used the CASA\footnote{\url{https://casa.nrao.edu/}} \citep{McMullin07}
software to extract the spectra in the position-position-velocity cubes 
for the (1,1) and (2,2) $\ammonia$  transitions to the contours defined
with the dendrograms. For these objects, the one-dimensional velocity 
dispersion was directly computed as $\veldisp = \mathrm{FWHM}/2 
\sqrt{2\ln2}$, where FWHM is the full width at half maximum obtained 
from the fit to the hyperfine structure of ammonia, performed with the 
``$\ammonia$(1,1)" CLASS method within the GILDAS\footnote{\url{http://
www.iram.fr/IRAMFR/GILDAS}} software. Table \ref{tab:g14} shows the 
properties derived for the sample in G14.  For this case we obtain objects 
in a size range of 0.05-0.5 pc and mean density $\sim 2\times10^4\pcc$ 
(see Fig. \ref{fig:histong14}).

\begin{figure}[]
\centering
\includegraphics[scale=0.54]{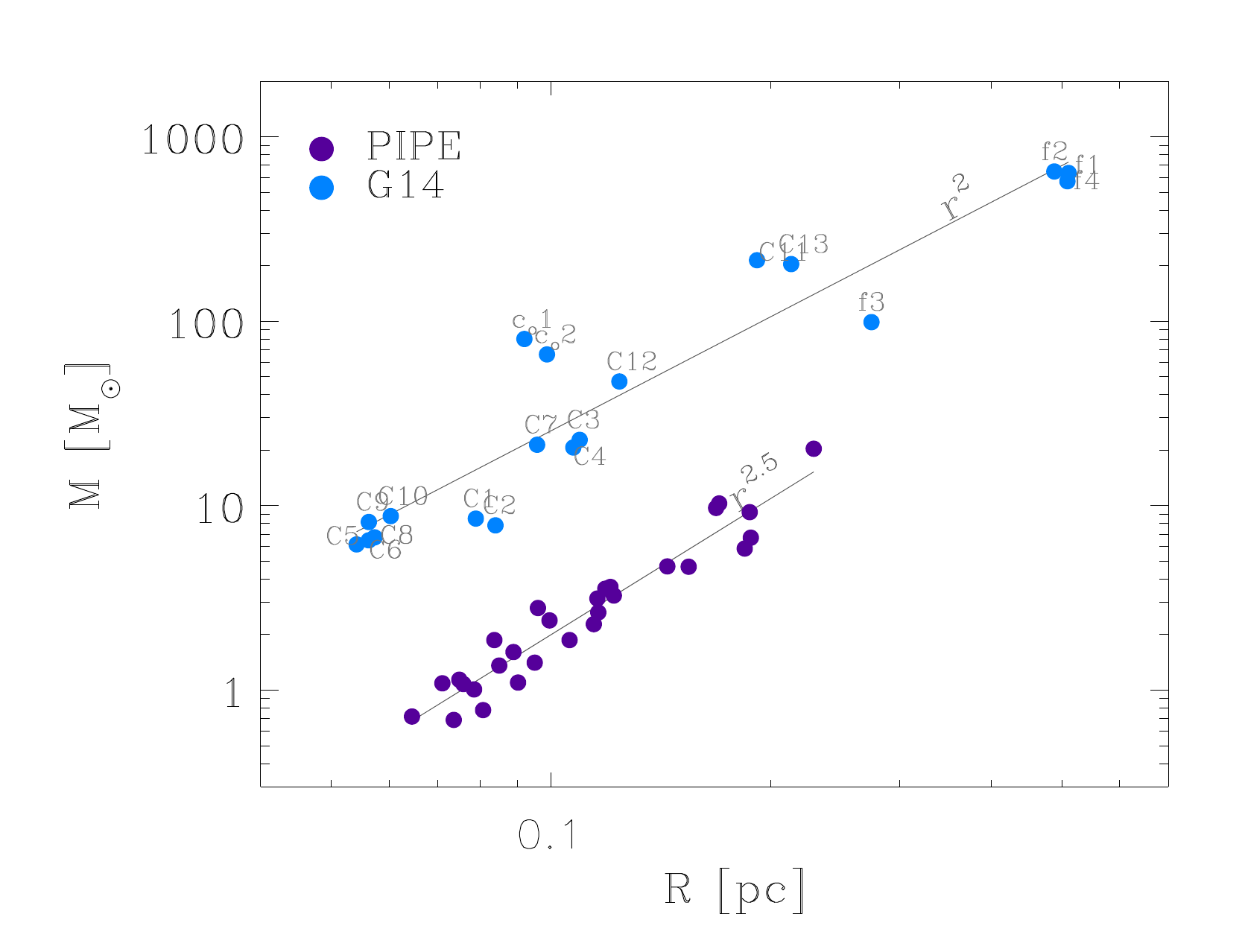} 
\caption{Size-mass relation for the selected sample of the Pipe cloud
(purple) and the G14 cloud (blue). The objects in the G14 sample 
correspond to the contours in Fig. \ref{fig:g14sample}. We additionally 
show the corresponding fits, $r^p$, for each sample, with $p \sim 2.5$ 
for the Pipe and $p \sim 2$ for the G14 clouds.}
\label{fig:obsporperties}
\end{figure}

Note that, in principle, we could derive the mass of the clumps in G14
directly from the line data for the (1,1) and (2,2) $\ammonia$ transitions,
assuming a certain ammonia abundance.  However, for consistency with 
the Pipe core sample, whose mass estimate was derived 
with the dust extinction map, we considered the column density map from
\citet{Lin17} to compute the masses within the contours defined in the 
ammonia maps with the \dendro\ algorithm, as described above, assuming 
that the two sets of observations trace the same gas.
We show in 
Appendix \ref{App:ammoniadata} a comparison deriving the mass from 
the ammonia data. Figure \ref{fig:obsporperties} shows the mass and size of 
the objects selected in both the Pipe and G14 clouds. On the other hand, 
Fig. \ref{fig:histong14} shows the density histogram for both clouds. We can 
observe that the Pipe cloud sample contains cores with similar densities, 
while the G14 sample covers a larger density range, varying from filaments 
to dense cores.

In this way, for both clouds we have the mass derived from dust emission 
and the velocity dispersion derived from the $\ammonia$ emission. 
Using the values of $M, R,$ and $\veldisp$ thus derived, we compute 
the virial parameter (eq.\ \ref{eq:alpha}) and Larson ratio (eq.\ \ref{eq:larsonr}) 
for the G14 clumps and cores that will be discussed in section \ref{s:results}. 
For the Pipe cores, we use the $M, R,$ and $\veldisp$ reported in the literature 
(Table \ref{tab:pipe}).

\begin{figure}[h]
\centering
\includegraphics[scale=0.53]{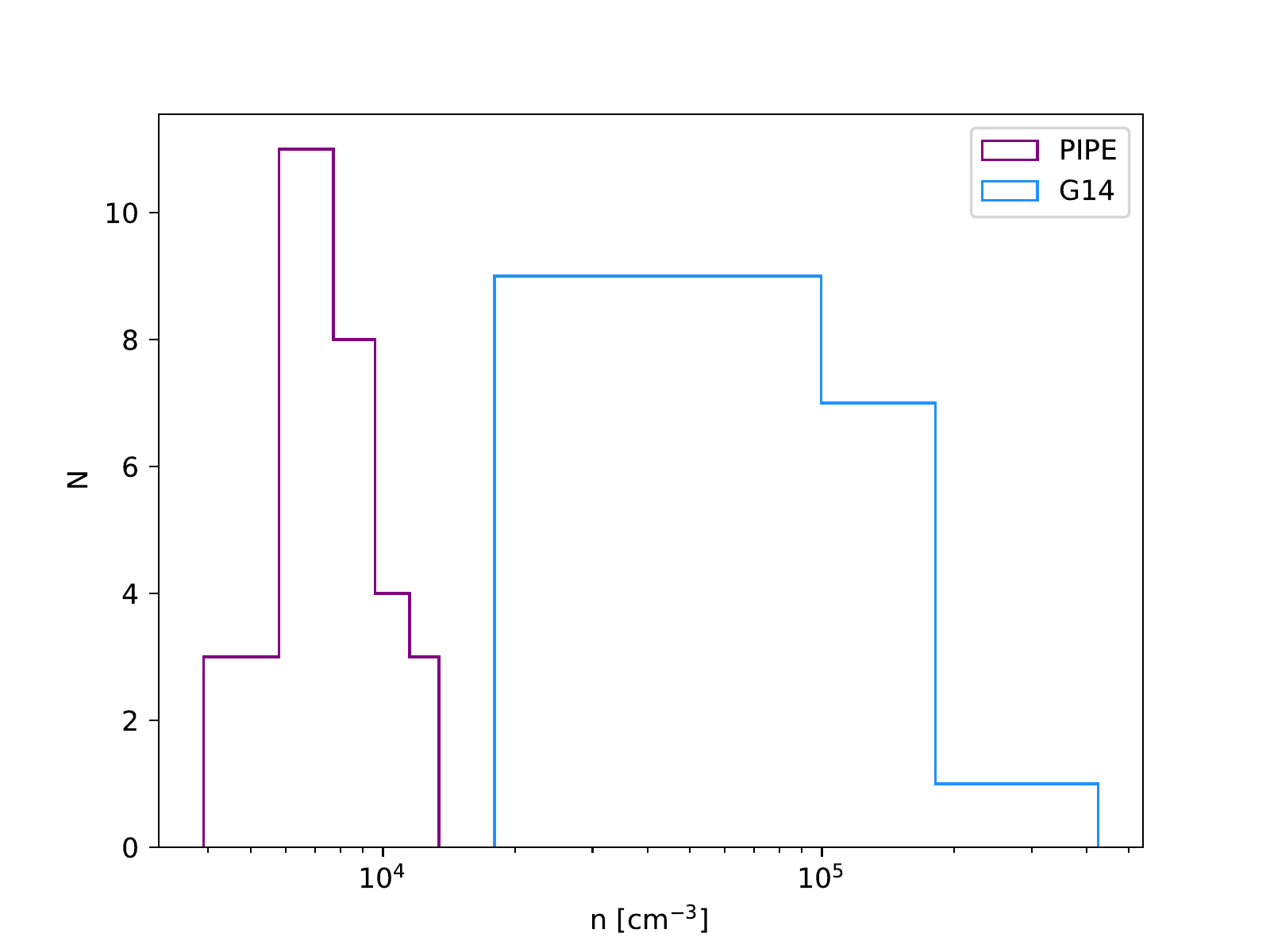} 
\caption{Density histograms for the selected sample of the Pipe cloud 
(purple) and the G14 cloud (blue).}
\label{fig:histong14}
\end{figure}

\subsection{The numerical sample} 
\label{subsec:numericalsample}

\begin{figure*}[ht!]
    \centering
        \includegraphics[width=17.cm]{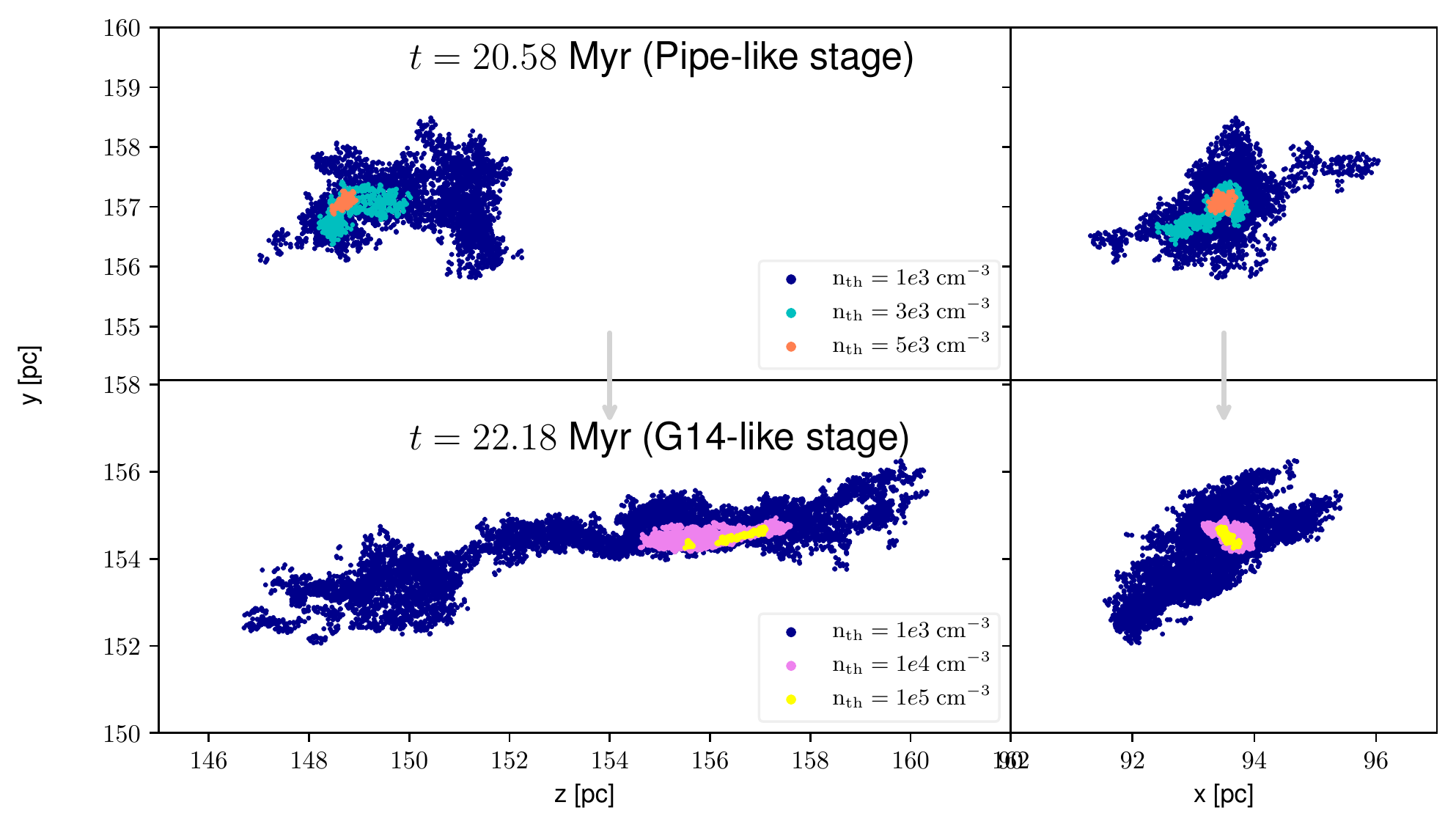}
    \caption{ Numerical clump at the initial and final times in two different
     planes and at different density thresholds. The top panels show the 
     numerical clump at $t=20.58$ Myr, the time when we define the 
     Pipe-like core (orange, $n_{th} =  5 \times 10^3 \pcc$). At this stage 
     the clump can be compared to the substructure in the Pipe. 
     The bottom panels show the numerical clump at an advanced stage, 
     $t=22.18$ Myr. At this time, the cloud can be compared with the
      filamentary structure in G14.}
  \label{fig:simclump}
\end{figure*}

In the simulation, several star forming regions are formed from the 
collapse of the density fluctuations present in the clouds. As the parent  
cloud collapses, its mean Jeans mass decreases. Then, dense turbulent 
fluctuations collapse once they exceed their Jeans mass and quickly start  
forming stellar particles \citep{Girichidis14, Zamora14, VS19}. Thus, most 
of them reach a high star formation activity in a short time. At these stages, 
feedback should  have important effects, affecting the kinematics around 
the clumps and reducing the SFR if it were included. However, it is not 
considered in the simulation. Thus, to avoid objects that should already be 
affected by feedback, possibly in the process of dispersal, we search for 
a star forming region in an intermediate time interval in the simulation 
($\approx 18-23$ Myr) such that the simulation has dissipated enough 
kinetic energy from the initial turbulent fluctuations in order to have realistic 
velocity dispersion values, and does not yet have too high a star formation 
activity at late stages, such that stellar feedback has not had time to 
affect the region (see Appendix \ref{app:feedback}).

\begin{figure*}[ht]
    \centering
           \includegraphics[width=8.5cm ]{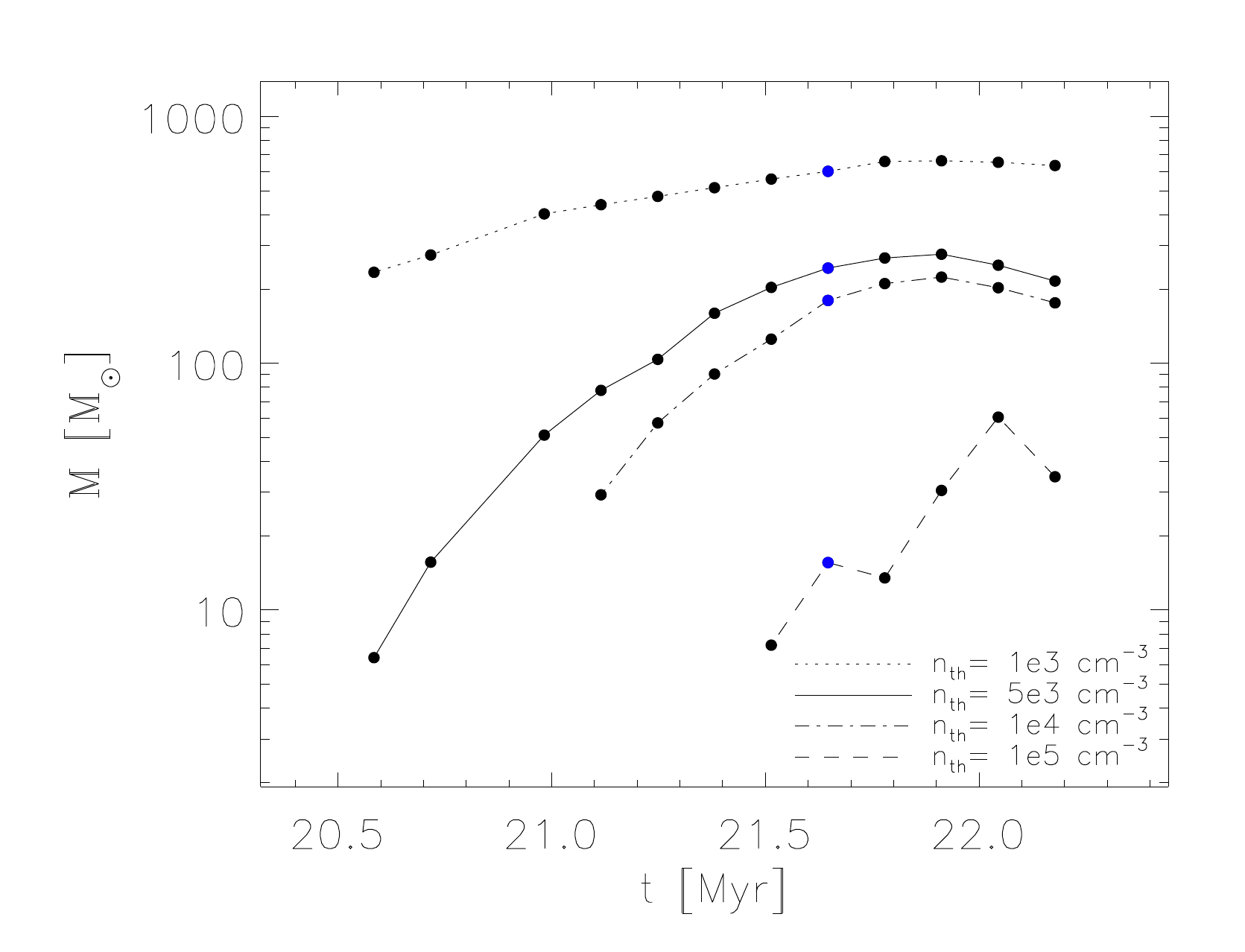} 
           \includegraphics[width=8.5cm ]{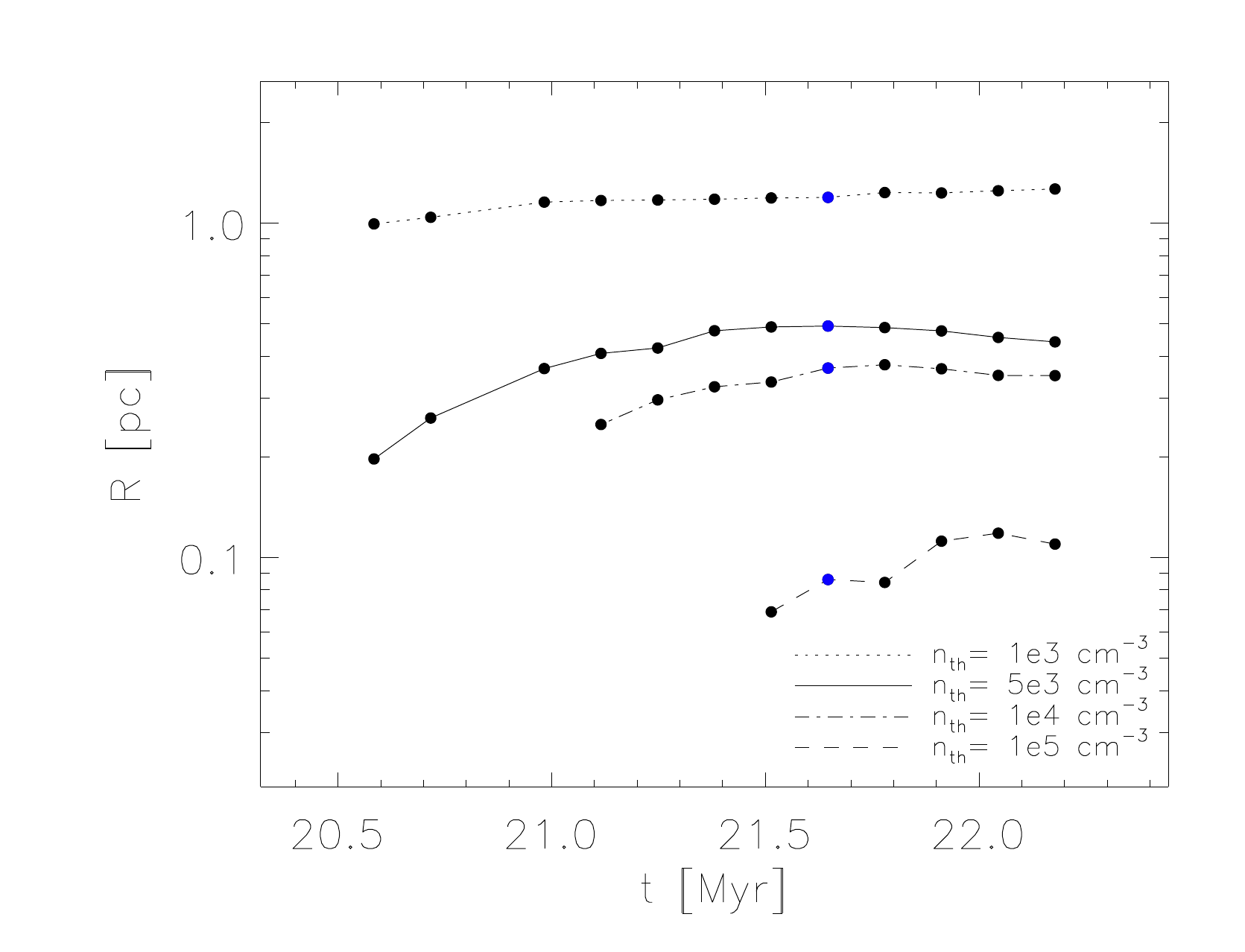}
    \caption{ Time evolution of the mass and size for the numerical clump. 
    The blue dot corresponds to the time when star formation begins, 
    while different lines correspond to the density thresholds $\nth$ used 
    to define substructures within the numerical clump.}
  \label{fig:mrt}
\end{figure*}

To define different categories of objects like filamentary clumps, hubs 
and dense cores, we consider density thresholds similar to the mean 
densities of the corresponding objects in the Pipe and the G14 clouds 
by applying the clump-finding algorithm 
for SPH particles described in \citet{Camacho16}\footnote{The clump-
finding algorithm \citep[see ][]{Camacho16} works by defining connected 
regions above a certain density threshold, $\nth$, in the density field. First, 
$\nth$ is set, and then the algorithm searches for the particle with the 
highest density above the threshold and searches for all the neighboring 
particles around it with densities larger than $\nth$. In order to have 
enough mass resolution we considered a minimum of 80 SPH particles in 
each clump. This algorithm has been designed for working directly on the 
SPH data and works similarly to the dendrograms algorithm used for the 
observations, except that it doesn't automatically track the lineage of the 
substructures.}. 

Furthermore, for the object selection in the available period, we look for 
a cloud that meets the following conditions:
\textit{a)} to be a single coherent object in the 3D space and not just a 
random superposition in projection and to remain so over a few million 
years; 
\textit{b)} to show a not too high star formation efficiency, for the entire 
cloud we tolerate a SFE $\sim 20 \%$;  
\textit{c)} not to be located near the edges of the box to allow a 
straightforward identification of the objects without cross boundary 
complications. 
The requirements mentioned here restrict the number of objects that 
can be analyzed. Moreover, the cores evolve rapidly in comparison to 
the time interval between snapshots, $\Delta t=0.133$ Myr. Therefore, 
the cores become larger, denser and more massive within a few 
snapshots.

We selected from the simulation a clump that fulfills the above criteria. 
We refer to it simply as ``the numerical clump".  This clump was followed
over a timespan $\sim 1.6$ Myr, from 20.58 to 22.18 Myr,  in a region of
$7\times8\times19$ pc in which the clump evolves. In what follows, we 
refer to this region as ``Region A'' (Fig.\ \ref{fig:boxsim}). The initial time 
was selected as the time when cores above a threshold 
$n_{\rm th}= 5\times10^3 \pcc$ first appear 
in this region.\footnote{We use the mean density of the Pipe cores ($\sim 
7\times10^3 \pcc$) as the reference to set the lowest density thresholds.} 
During  the above timespan, the clump evolves from being similar to the 
substructure in the Pipe to being similar to the substructure in G14, and 
changes in shape, size, mass, and mean and maximum density 
(Figs.\ \ref{fig:simclump}-\ref{fig:mrt}). 
Note that, somewhat counterintuitevly, the clump \textit{grows} in mass 
and size during the process, due to the accretion of external material.

For the Pipe-like stage of the numerical clump, we 
recall that the Pipe cloud has dimensions $\sim 3$ pc $\times14$ pc, 
and that the more massive cores have masses from $\sim 5-20\ \Msun$, 
sizes of $\sim 0.2$ pc, and densities of $\sim 7\times10^3\pcc$. 
We thus define the numerical clump by a threshold $\nth=10^3\pcc$ 
and a Pipe-like core at $\nth= 5\times10^3 \pcc$. With these thresholds, 
the numerical clump has projected dimensions of $\sim 3\times5$ pc, 
and contains one roundish core of size $\sim 0.2$ pc, mass $ 6.3\ \Msun$,
and mean density $\sim 6330 \pcc$, in good agreement with a typical 
Pipe core. In order for this object to have dimensions comparable to those 
of the Pipe cloud, we need to go down to a threshold of $\nth= 300\pcc$, 
for which the resulting numerical cloud has dimensions $\sim 7\ {\rm pc} 
\times10 \ {\rm pc} \times15 \ {\rm pc}$, and mass $\sim 1100 \Msun$. 
This mass is still $\sim 8$ times lower than the Pipe's mass as reported 
by  \citet{Lada10}, over an extension of $3\times14$ pc. This suggests 
that the density profile at the Pipe stage is steeper than in the Pipe, 
perhaps due to the absence of magnetic field in our simulation, 
which could tend to smooth out the cloud somewhat.

For the G14-like stage, we defined the clump at $\nth= 10^3 \pcc$, 
denser regions at $\nth=10^4$, and a dense core at $\nth= 10^5 \pcc$.
As with the Pipe-like stage, we look for the structure in this epoch at 
$\nth=300\ \pcc$. The corresponding  mass and projected dimensions 
at this time are $\sim 1900\ \Msun$ and $\sim 9$ pc $\times25$ pc.
We see that the total mass of the numerical cloud at this threshold is still 
a factor of $\sim 10$ times lower than the mass of G14, while being longer 
and narrower. Thus the apparently steeper density profile of the numerical 
cloud remains at this stage.

For the  structures thus identified, we compute the mass as 
$M=nm\mathrm{_{sph}}$, where $n$ is the total number of particles 
belonging to the clump and $m\mathrm{_{sph}}$ is the mass of an SPH 
particle ($m\mathrm{_{sph}} = 0.06  \thinspace \Msun$); the size as 
$R=(3V/4\pi)^{1/3}$, where $V$ is the total volume of the 
particles\footnote{The total volume is the sum of the specific volumes of 
all particles, see \citet{Camacho16} for details.}; the column density as 
$\Sigma=M/\pi R^2$; the Larson ratio $\mathcal{L}=\sigma_{v,1D}/
R^{1/2}$; and the virial parameter. The latter is computed in the standard 
form as in eq. (\ref{eq:alpha}). We also estimated the instantaneous
star formation rate as SFR $ \approx \Delta \Mstar/2\Delta t$,  where 
$\Delta \Mstar$ is the mass that is transformed into sink particles during  
the time interval $2\Delta t$, and $\Delta t$  is the time between snapshots. 
The mass of the sink particles is not constant, since they accrete 
material from the surroundings as the simulation evolves. Thus, 
$ \Delta \Mstar$ includes the mass in new sinks as well as the mass
accreted by existing sinks. Finally, the star formation efficiency at any 
instant in time $t$ is computed as
${\rm SFE}(t) = \Mstar(t)/\left[M_{\rm gas}(t) + \Mstar(t)\right]$.


 \section{Results}\label{s:results} 

Observations from molecular clouds to dense cores in 
$\mathcal{L}-\Sigma$ diagrams have shown objects below or above the 
energy equipartition, and where the deviations have been explained 
assuming unbound clouds, external pressure confinement or turbulence 
regulation \citep{Heyer09, Field11, Traficante18a}. In the present work we 
offer a different interpretation for the sub-virial cores.

 \subsection{The numerical clump sample} 

\begin{figure}[ht!]
\centering
 \includegraphics[scale=0.5]{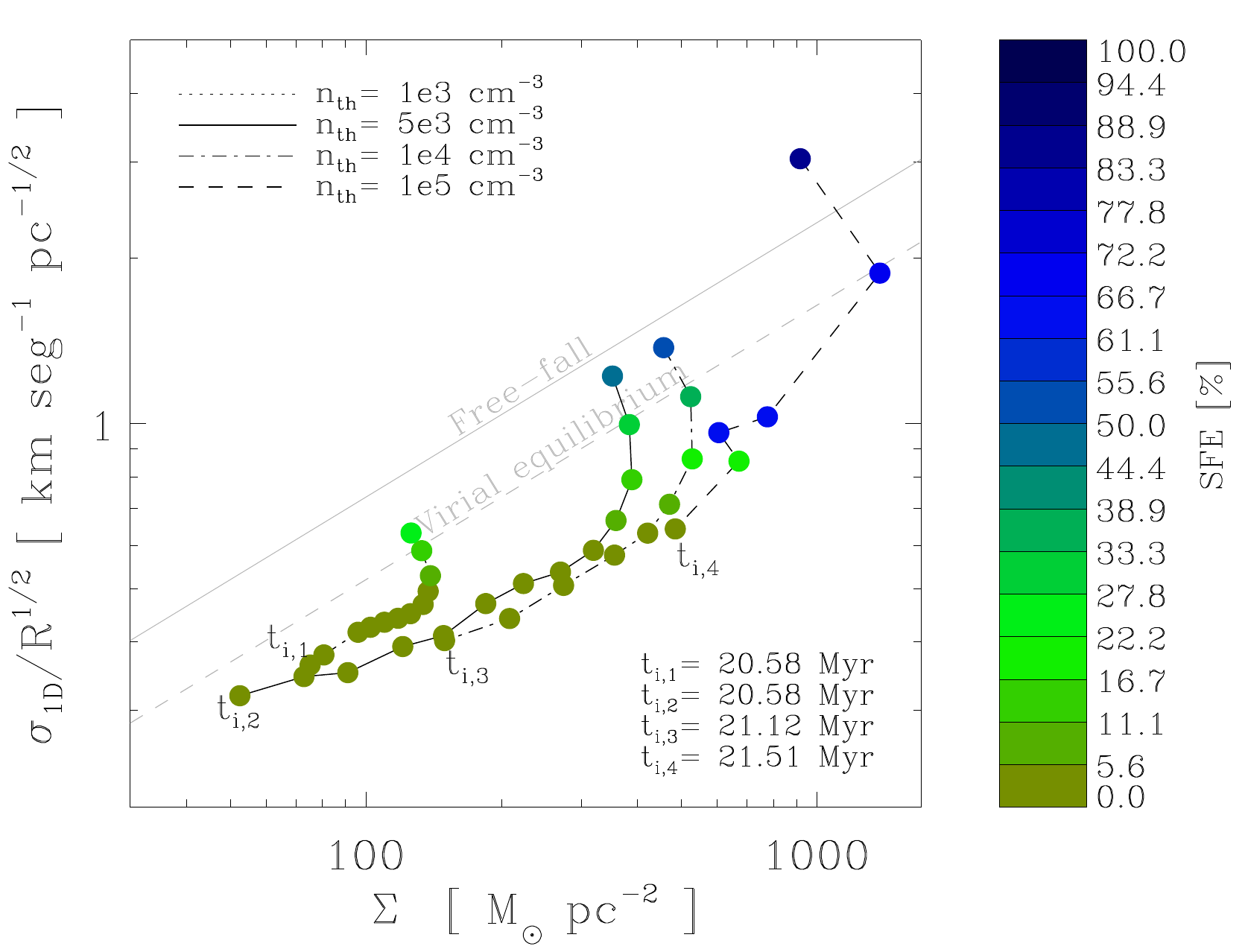}   
 \includegraphics[scale=0.5]{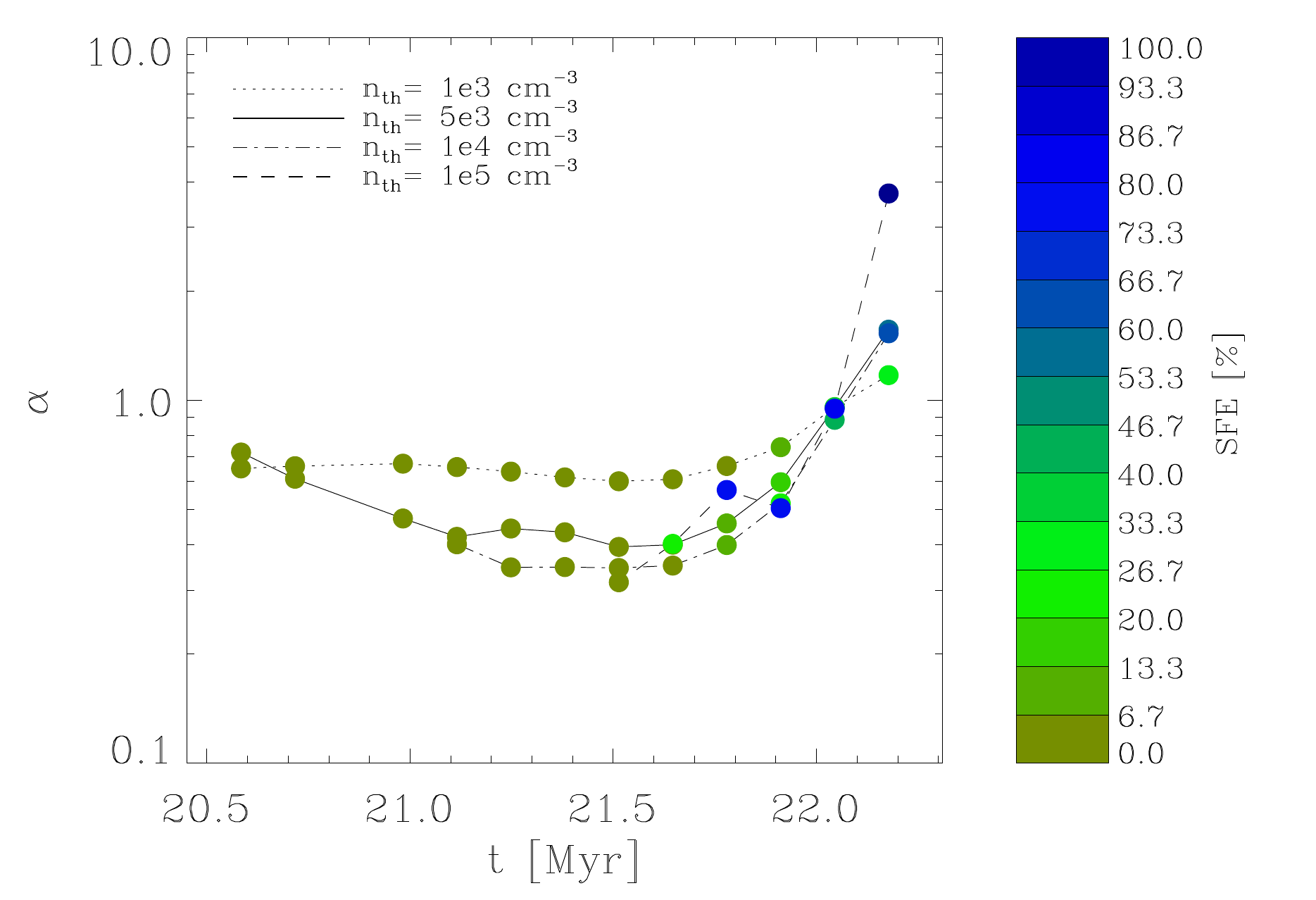} 
 \caption{Evolution of the substructures in the numerical clump in 
 the $\mathcal{L}-\Sigma$ diagram (top) and evolution of their virial
 parameter (bottom). Each point represents a time step in the simulation, 
 the time interval between them is $\sim 0.13$ Myr. The times when the 
 structures appear for the first time at each threshold are indicated by 
 $\rm{t_{i,n}}$. The final time is 22.18 Myr for structures defined at all 
 density thresholds. Also shown is the evolution of the SFE, indicated 
 by the colorbar.}
\label{fig:simsfe}
\end{figure}

\begin{figure}[h]
\centering
 \includegraphics[scale=0.5]{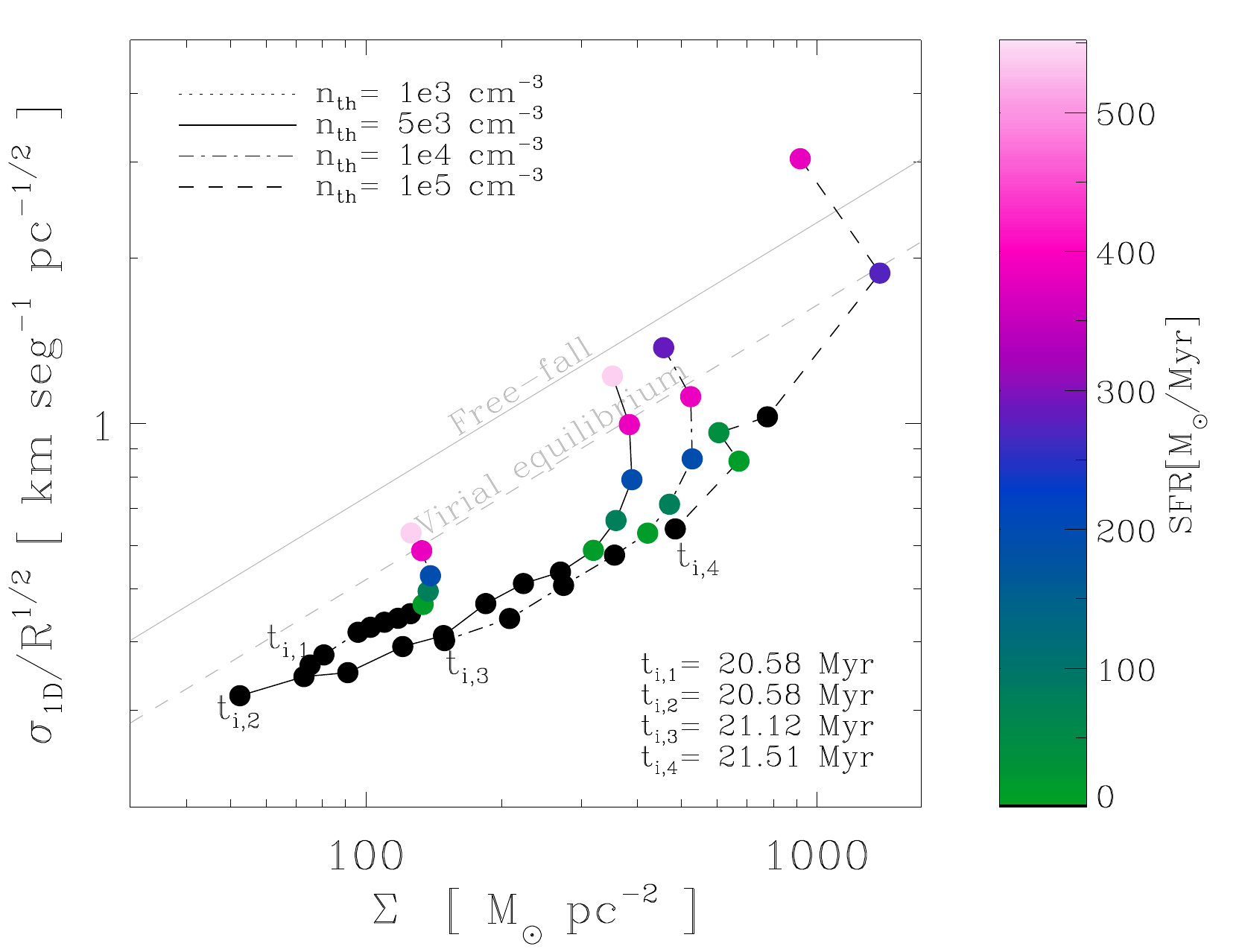} 
 \includegraphics[scale=0.5]{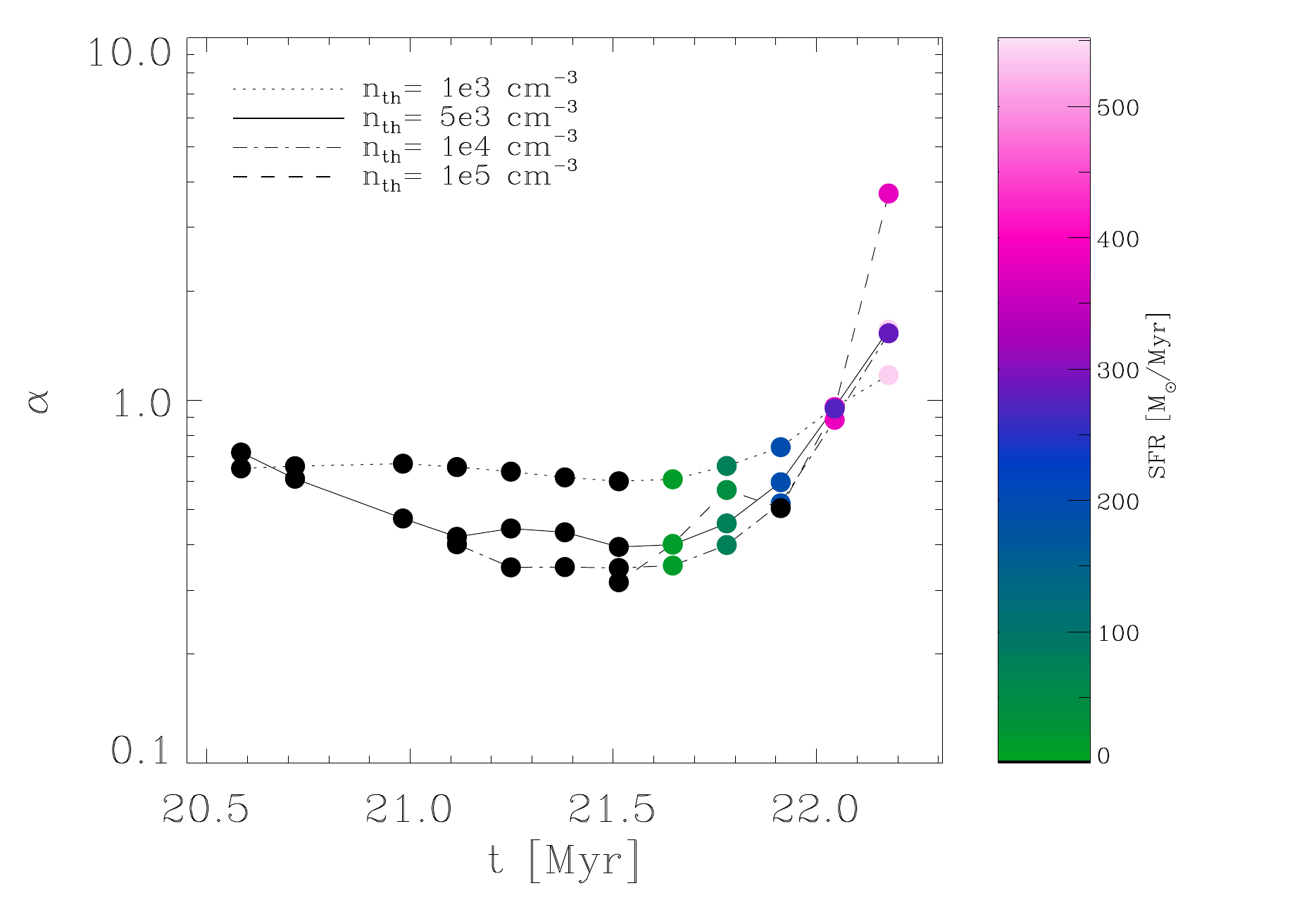}
  \caption{ Same as Fig. \ref{fig:simsfe} although in these plots the 
  color scheme indicates the SFR.}
  \label{fig:simsfr}
\end{figure}

The energy balance in the numerical clump sample is studied in the 
$\mathcal{L}-\Sigma$ diagram (top panels of Figs.\ \ref{fig:simsfe}
and \ref{fig:simsfr}. The evolution of the virial parameter $\alpha$ is 
shown in the bottom panels. In addition, in these figures the symbols 
are colored according to the SFE (Fig.\ \ref{fig:simsfe}) and the SFR 
(Fig.\ \ref{fig:simsfr}). In the $\mathcal{L}-\Sigma$ diagram, we refer 
to the line $\mathcal{L}=\sqrt{G\Sigma}$ as the ``virial line", and to
the line $\mathcal{L}=\sqrt{2G\Sigma}$ as the ``free-fall line" 
\citep{Ballesteros11}. Since these two lines are so close to each other, 
we collectively refer to both as ``the equipartition condition". Note that 
they correspond to values of the virial parameter of $\alpha=1$ and 
$\alpha=2$, respectively. We refer to the region above energy 
equipartition ``super-virial'' and to the region below, as ``sub-virial''. 

The various lines correspond to the density thresholds used to define the 
substructure in the numerical clump, joining objects defined by the same 
density threshold. These  can be thought of as objects seen in tracers with 
different critical densities. 
Thus, the time evolution for an object at a certain 
density can be seen by following the connected symbols from left to right.
We observe in the simulation that denser objects, and eventually stars, 
appear later in the evolution of the clump. Because of this, we label the 
beginning of each curve with the time $t_{i,n}$ when the objects first apear 
at $\nth$.
In fact, it is noteworthy that the first stars appear in the clump 
approximately one free-fall time  of the clump after the time when we first 
observe it. Indeed, at that time ($t_0= 20.58$ Myr), the clump's mean 
density is $\rho = 2.35\times10^3\ \pcc$, for which the free-fall time 
$t_{\rm ff}=\sqrt{3\pi/ (32 G\rho)}$ is $ \approx 0.95$ Myr. On the other 
hand, the first sink appears $1.06$ Myr  after $t_0$, in very good 
agreement with the value of $t_{\rm ff}$, considering that the actual 
collapse is always slightly slower because thermal pressure is not totally 
negligible \citep{Larson69}. This shows that the clump is evolving 
essentially under the action of its own self-gravity. 

There are some important features to notice in the 
$\mathcal{L}-\Sigma$ diagrams: 
\textit{i)} the earliest structures appear sub-virial for both low and high 
densities;  \textit{ii)} as the objects evolve in time (see also $\alpha\ vs. \ t$ 
plots), they approach equipartition and in one case the object even 
becomes super-virial. This can be seen for objects defined at all density 
thresholds.

Additionally, a gradual increase is seen of both the SFR and the SFE, 
represented by the colors in Figs.\ \ref{fig:simsfe} and \ref{fig:simsfr}. 
That is, an increase of the SFE and SFR occurs simultaneously with 
the increase of kinetic energy, manifested in the variation of the Larson 
ratio and the virial parameter.
The increasing star formation activity is a natural consequence for a clump 
that becomes denser on average due to global gravitational contraction 
and thus contains a higher fraction of high-density gas 
\citep{Zamora14,VS18} which is responsible for the `instantaneous' star 
formation in the clump.  Thus, {\it we suggest that SFR and SFE evolve 
simultaneously with the energy budget of the clumps}.

It is observed that, early in their evolution, the objects start with low 
values of $\alpha$; in fact, the clump defined at $\nth=10^3\pcc$ 
remains with $\alpha$ roughly constant over $\sim 1$ Myr. For some 
other clumps, the virial parameter  even decreases before star 
formation begins, to later increase again, approaching the free-fall 
value ($\alpha=2$) at times when the SFR has reached values of a
few $\times 100\, \Msun$ Myr$^{-1}$. This shows that values of 
$1 \lesssim \alpha \lesssim 2$ are not necessarily a signature of 
unbound objects. Instead, they may be indicative of the  approach 
to the free-fall value.

\begin{figure}[h!]
\centering
\includegraphics[scale=0.5]{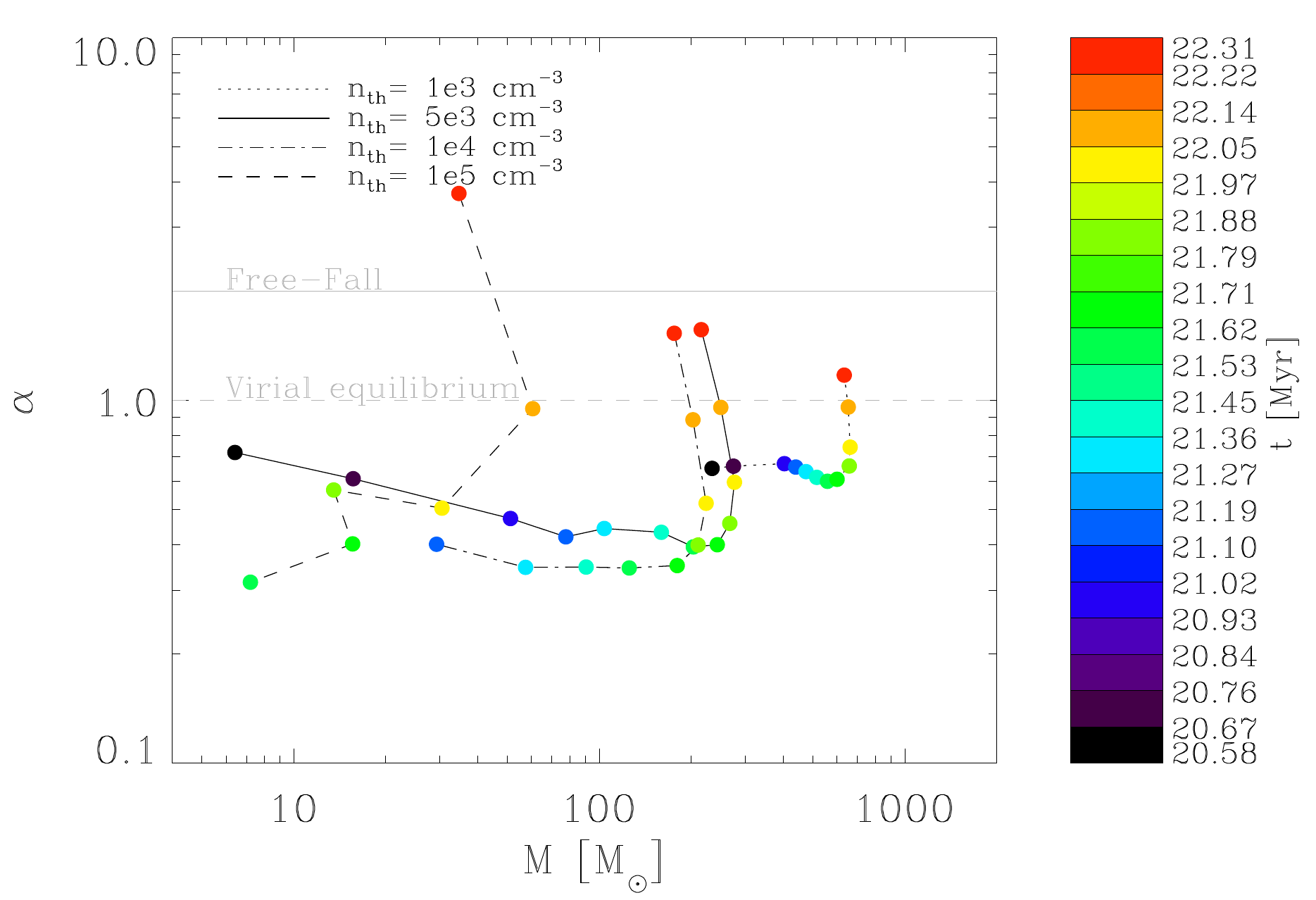} 
\caption{Virial parameter $\alpha$ as a function of mass for the 
numerical clump and its substructureS. The time sequence is 
shown in colors, while the lines show the density thresholds.}  
\label{fig:alphaclump}
\end{figure}

\begin{figure}[ht!]
\centering
 \includegraphics[scale=0.5]{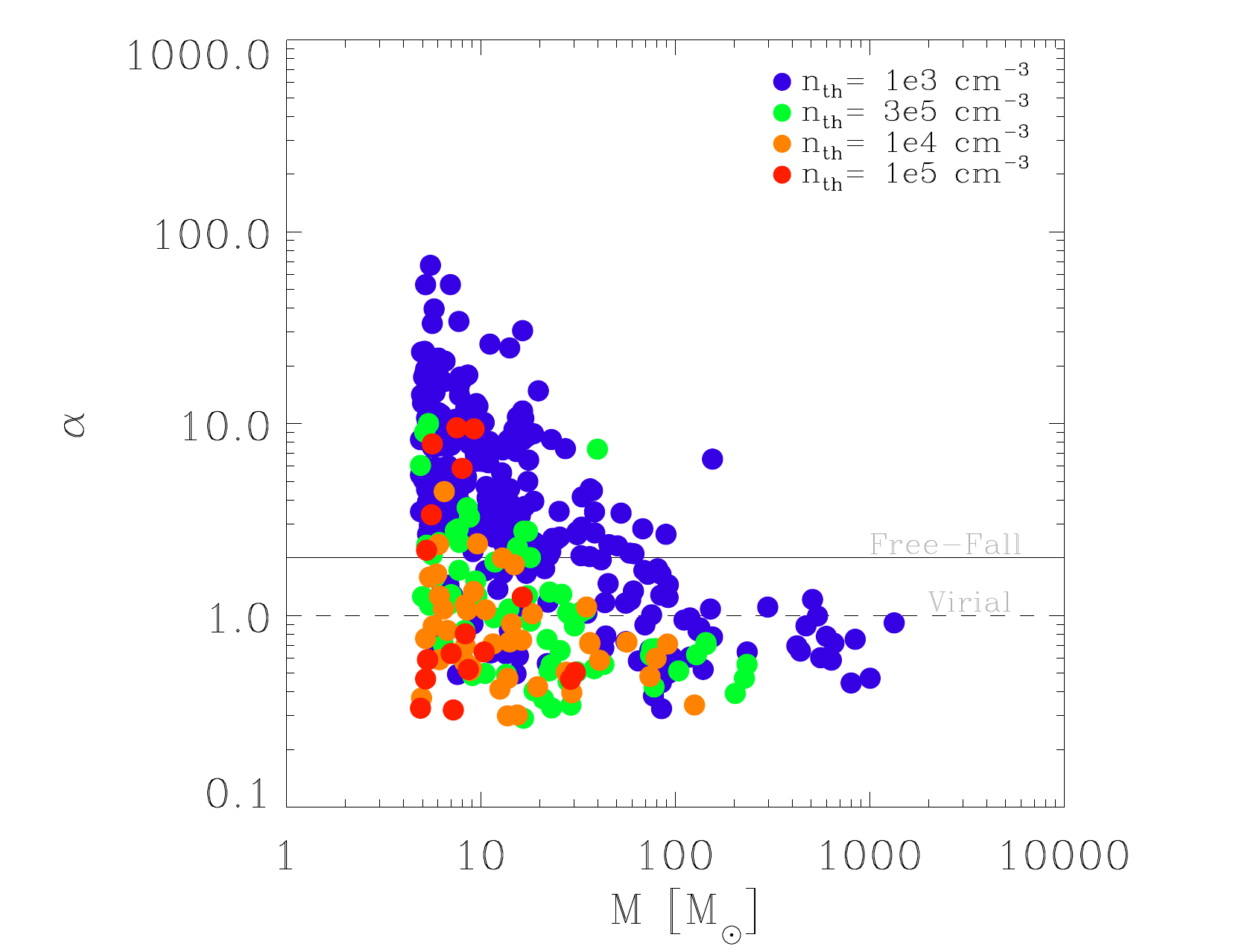}  
\caption{Virial parameter for a sample of clumps from the
RUN03 simulation. We considered $t=\ 20.8,\ 21.2,\ 22.2,\ 24.8$ 
and 26.5 Myr and the same density thresholds for the numerical 
clump.}  
\label{fig:alphasim}
\end{figure}

In addition, Fig. \ref{fig:alphaclump} shows the evolution of the 
numerical clump's substructure in the $\alpha$-$M$ diagram. 
As in Figs.\ \ref{fig:simsfe} and \ref{fig:simsfr}, the lines connect 
clumps defined at the same density threshold at different times, 
indicated by the colorbar. Thus, the set of points shown with a 
particular color illustrates the hierarchy of nested structures at 
a given time, while points joined by each line represent the 
evolution of a clump defined at that  threshold.  It is noteworthy 
that the slope of the clump hierarchy at the last time exhibits a 
negative value. 
This implies that the most massive objects in a 
statistical sample, and in a sample including different density 
thresholds and for later stages, have lower values of $\alpha$ 
than the less massive ones, as shown in Fig.\ \ref{fig:alphasim}, 
in spite of the fact that each  structure is seen to evolve from 
sub-virialization to equipartition, as predicted by eq.\ 
(\ref{eq:alpha-Sigma}). This is because, as shown in \citet{VS19}, 
for a coeval sample of cores selected in such a way that their mass 
scales as $R^p$, with $p <3$, as is often the case (see Fig.
\ref{fig:obsporperties}), the more massive objects have lower 
densities, and therefore longer free-fall times. This implies that, 
at some age $t$, the more massive objects have traversed a 
smaller fraction of their free-fall time, and are therefore at earlier 
stages of their own evolution, therefore being more sub-virial.

\begin{figure*}[ht!]
\centering
\includegraphics[scale=0.7]{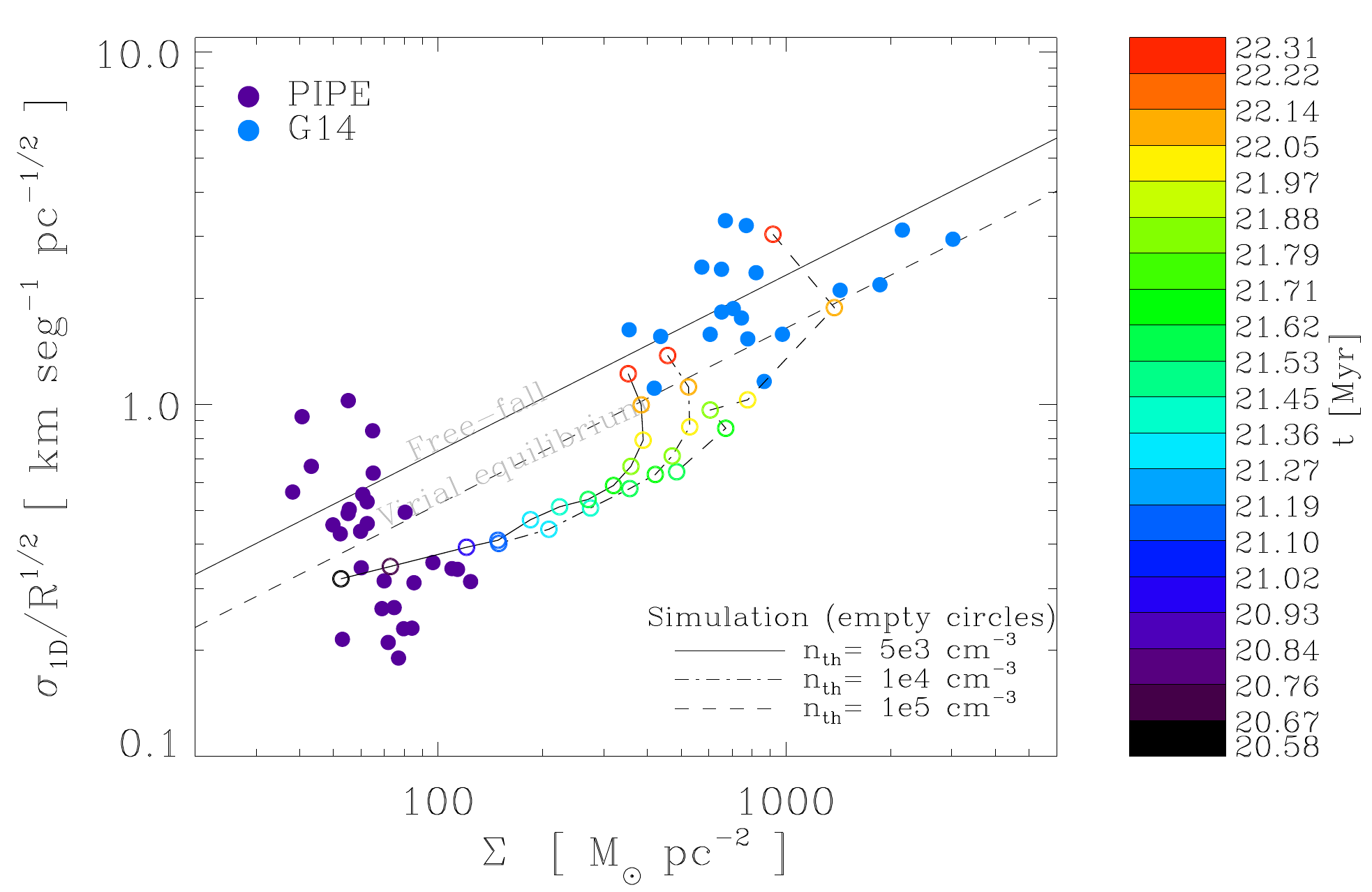}
\caption{$\mathcal{L}-\Sigma$ diagram for the observational clump 
sample (purple and blue dots) and comparison with the temporal 
evolution for some of the substructure in the numerical clump. 
The simulated clumps (empty circles) evolve from the left to the right, 
their colors correspond to the color bar in Fig. \ref{fig:alphaclump}, 
which represents time.} 
\label{fig:khobs}
\end{figure*}	

 \subsection{Comparison with the observational sample} 
 \label{sec:resultobs}

As mentioned in Sec. \ref{s:data}, the two real clouds considered 
in this work are  expected to be in different evolutionary stages.  
Indeed, the Pipe shows little signs of star formation \citep[the component 
known as B59 or the Mouthpiece is known to have embedded a group of 
young stars, e.g.,][]{Brooke07, Roman10, Dzib13} and a low fraction 
of its mass at high density \citep{Lada10}, which has been interpreted as 
an indication of an early evolutionary stage \citep[e.g.,][]{Onishi99,
Rathborne08, Lada10,Frau10, Frau15,VS18}. On the other hand, G14  
exhibits active star formation and a duration of the star formation activity 
of a few Myr, suggestive of a somewhat more advanced evolutionary 
stage \citep{Povich16}. 

\begin{figure*}[ht!]
\centering
\includegraphics[scale=0.7]{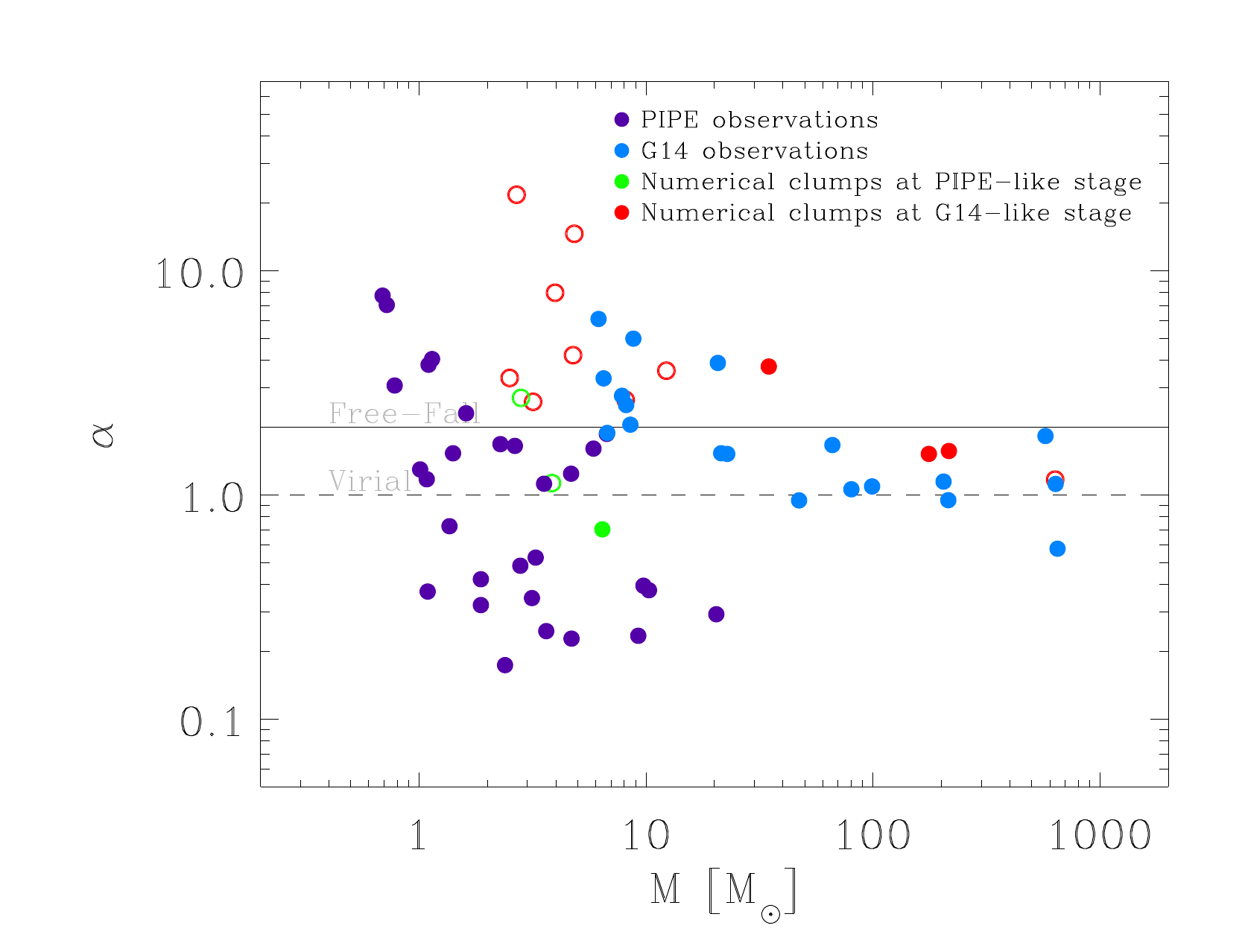}
\caption{Virial parameter for the Pipe (purple dots) and  G14 
(blue dots) selected clumps. Additionally, other objects in Region 
A but outside the numerical clump are shown at the Pipe-like stage 
($t= 20.58$ Myr and $\nth = 5\times10^3 \pcc$) and at the G14-like 
stage ($t=22.18$ Myr), in green and red empty circles respectively. 
The numerical clump and its substructures are denoted by the filled
circles, with green and red again corresponding to the Pipe-like and 
G14-like stages, respectively.}
\label{fig:obs}
\end{figure*}

Similarly to what was done for the numerical sample, we plot the 
corresponding $\mathcal{L}-\Sigma$ diagram (Fig.\ \ref{fig:khobs}) and 
$\alpha$-mass plot (Fig.\ \ref{fig:obs}) for the substructures in the Pipe
(purple dots) and G14 (blue dots) clouds. 
In Fig. \ref{fig:khobs}, we plot the evolution of the substructures of the
numerical clump defined at $\nth = 5\times10^3, 10^4$ and $10^5\ \pcc$, 
with colors representing the time evolution with the same scheme as 
in Fig.\ \ref{fig:alphaclump}. In particular, the clump at $5\times 10^{3}
\pcc$ (black circle) named the ``Pipe-like core"  
(Sec. \ref{subsec:numericalsample}), evolves from a locus centered in 
that of the Pipe cores to one close to the G14 sample. When the 
Pipe-like core first appears (at 20.58 Myr),  it shows a structure similar 
to the dense cores in the Pipe cloud with, $M\sim 6.4\ \Msun,\ R \sim 0.2$ 
pc and $\sigma_{1D}\sim 0.14\  \kms$, and exhibits a low virial parameter 
$\sim 0.7$. As the simulation evolves, the object defined by the same 
threshold as the Pipe-like core grows in size and mass due to accretion, 
changing its morphology from roughly spherical to a more elongated 
structure (its largest axis has $\sim 3$ pc)\footnote{Recall that the size 
definition considered in this work assumes that all particles belonging to 
a clump are contained in a sphere whose volume is the sum of all the 
particles' volumes. Thus, the size, $R=(3V/4\pi)^{1/3}$, does not reflect 
the largest dimension of the clumps.}. Its mass grows in time even after 
the star formation begins (see top panel in Fig.\ \ref{fig:mrt}). However, 
after $\sim 0.3$ Myr, it becomes roughly constant.

At the latest time, the clump defined at $10^3\ \pcc$ (Fig.\ \ref{fig:mrt}), 
which has a filamentary shape (Fig.\ \ref{fig:simclump}) can be compared 
with the filaments f1 and f2 in G14 (see Tab.\ \ref{tab:g14}).
In the simulation this filamentary clump has mass $\sim 600\ \Msun$ and 
projected dimensions $\sim 4 \times 13$ pc, similar to the filaments in G14. 
On the other hand, the interior regions of the Pipe-like core within 
higher-$\nth$ thresholds (yellow and red circles in Fig.\ \ref{fig:khobs}) 
evolve to positions comparable to those of denser G14 clumps and 
filaments. At time $t=$ 22.04 Myr, the yellow and red circles at 
$\nth = 10^4 \pcc$ have $M \sim 203 \ \Msun, \ R\ \sim 0.4 \ \rm{pc},\ 
\sigma_{1D} \sim 0.6\ \kms$, and a virial parameter $\sim 0.9$, between 
f1 and f3 of the G14 sample, while the yellow circle at $\nth = 10^5 \pcc$ 
has $M \sim 60 \ \Msun,\ R \sim 0.1 \ \rm{pc}, \sigma_{1D}\sim 0.6\ \kms,$
and a virial parameter $\sim 0.9$, in the neighborhood of C11, C12, and
C13 (see Tab.\ \ref{tab:g14}).

We observe that roughy half of the cores in the Pipe sample (which 
we consider less evolved; see Sec. \ref{s:data}) appear  in the 
super-virial range in the $\mathcal{L}-\Sigma$ and the $\alpha-M$
diagrams. In \citet{Camacho16} we show that the scatter around 
equipartition for objects in the low-$\Sigma$ range might be caused 
either by the dispersion of the clumps or by large scale motions 
assembling them. In the present study, we observe that the super-virial 
Pipe cores, in Figs.\ \ref{fig:khobs} and \ref{fig:obs}, are the smallest 
and less massive objects.  We suggest that, as discussed in 
\citet{Camacho16}, in about half the cases, their apparent kinetic energy 
excess may be a consequence of the externally-driven motions that are 
assembling them, rather than a signature of being unbound. In the other 
half, actual dispersal may be occurring, but in no case the objects are 
confined by large external thermal pressure. On the other hand, the 
sub-virial cores are in fact the most massive ones, and include some of 
the cores belonging to the active star forming region B59. As discussed 
in \citet{Ballesteros18} and \citet{VS19}, their motions may be dominated 
by their self-gravity, although without having reached the free-fall velocity 
corresponding to equipartition (cf.\ eq.\ (\ref{eq:Lcal-Sigma})). In contrast, 
the G14 sample (which we consider more evolved) is seen to lack strongly 
sub-virial objects. This is consistent  with the evolution of the numerical 
sample from sub-virial to super-virial states.

To compare with the numerical sample, in Fig.\ \ref{fig:obs} we show the 
numerical clump and its substructures, at both the Pipe-like (green filled 
circles) and the G14-like (red filled circles) stages, 20.58 and 22.18 Myr 
respectively. Furthermore, we also show other clumps and cores located 
within Region A but outside the numerical clump, shown with green and 
red empty circles for the Pipe-like and the G14-like stages respectively. 
Note that in order to include these, we had to reduce our mass resolution 
criterion to 40 SPH particles, since no additional objects other than the 
numerical clump with more than 80 SPH particles are found within Region 
A in this time interval. We can observe that the numerical samples, at their 
corresponding stages, are consistent with the real clumps, occupying 
similar loci and exhibiting a negative slope in the $\alpha$-$M$ diagram, 
with the more massive objects having lower $\alpha$ values.

\begin{figure*}[]
\centering
\includegraphics[angle=270,scale=0.6]{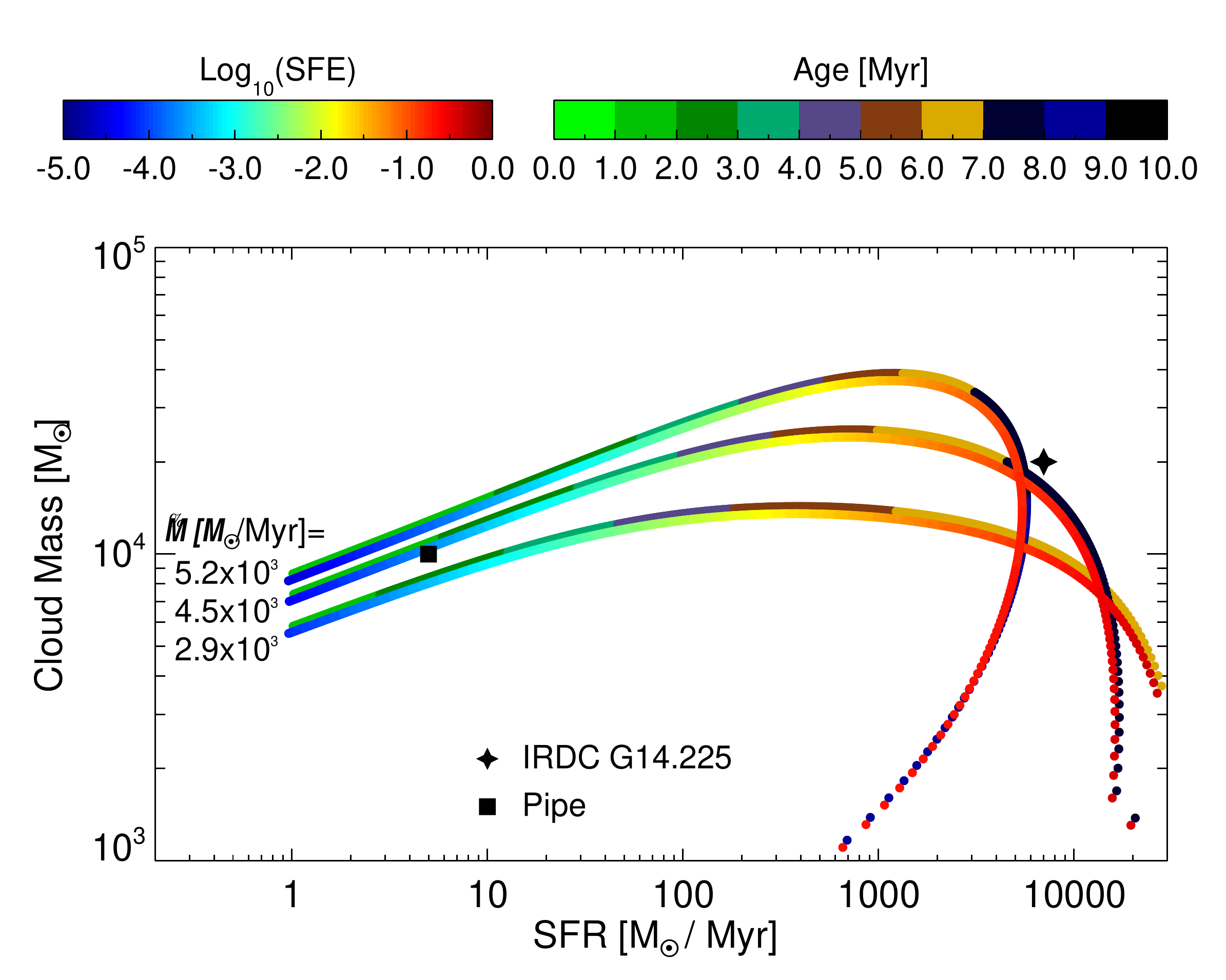}
\caption{Analytical model for the  evolution of clouds with different 
accretion rates in a cloud mass {\it versus} SFR diagram \citep{VS18}. 
The tracks show the instantaneous mass and SFR for the different 
accretion rates. The symbols correspond to the loci of the Pipe and G14 
clouds in this diagram. Each track is shown in two color scales 
representing the cloud age and its SFE. Note that at $\sim$ 6 Myr all 
models reach its maximum mass and then the cloud mass decreases 
and the SFR decelerates due mainly to effects of ionizing radiation from 
massive stars. }
\label{fig:zamoramodel}
\end{figure*}

Concerning the star formation activity, \citet{Lada10} report an SFR 
$\sim 5\, \Msun$ Myr$^{-1}$ for the Pipe, while \citet{Povich16} report 
an SFR $\sim 7 \times 10^3\, \Msun$ Myr$^{-1}$ for G14, or roughly 
3 orders of magnitude larger than that of the Pipe. Of course, a more
meaningful comparison is in terms of the {\it specific} SFR, or {\it sfr}, 
since G14 is clearly a more massive region than the Pipe. \citet{Lada10} 
report a total mass for the Pipe of $\sim 8 \times 10^3\, \Msun$, while 
\citet{Elmegreen79} quote a mass $\sim 2 \times 10^5\, \Msun$ 
for  M17 S Wex \citep[see also, ][]{Povich16, Lin17}, the parent cloud 
of the G14 IRDC. Thus, {\it sfr} $\sim 6.2 \times10^{-4}$ Myr$^{-1}$ for 
the Pipe, while {\it sfr} $\sim 3.5 \times 10^{-2}$ Myr$^{-1}$. 
Thus, the specific SFR of G14.2 is $\sim 50 \times$ larger than that of 
the Pipe, implying indeed a much stronger star-forming activity even 
when normalized to the total cloud mass. Additionally, a recent study 
by \citet{Shimoikura19} reported a SFE $9-17\%$ for the M17 S Wex. 
This study, conducted on N$_2$H$^+$, concluded that according to 
the density profile of the cores, more than 80\% of their sample is 
consistent with the free-fall condition. This result reinforced our 
evolutionary scenario in terms of the gravitational collapse.

\subsection{Comparison with an analytical model}

We use the model presented by \citet[][see also Zamora-Avil\'es \&
V\'azquez-Semadeni, 2014]{Zamora12}, which describes the evolution 
of clouds with different masses, to  compute the evolution for G14, as
shown for several clouds in \citet{VS18}. The model starts with a cloud 
formed by converging streams in the warm neutral medium (WNM). 
The cloud mass grows due to accretion at constant density until it 
reaches its Jeans mass and starts to collapse. The collapse is solved 
self-consistently during the evolution of the cloud. As this happens, 
the mean density  and mass fraction at high density increase, implying 
an increase of the SFR. Thus, the model is able to describe, as a function 
of time, and parameterized by the maximum mass attained by the cloud 
during its evolution, the SFR, SFE, mass and mean density of clouds in 
global gravitational contraction.  
The model predicts that, during the early stages of a cloud's evolution, 
the mass, density and SFR increase monotonically, while the radius 
decreases. Therefore, during this period, any of these quantities can be 
used as a proxy for time. For the present work, the model has been 
parameterized by the accretion rate onto the cloud. We consider models 
with mass accretion rates  $\dot M = 2.9, 4.5, 5.2 \times 10^3\ \Msun$
Myr$^{-1}$, in order to increase the mass of the cloud by amounts 
comparable to the mass difference between G14 and the Pipe. Figure 
\ref{fig:zamoramodel} shows a diagram of instantaneous cloud mass vs.\ 
SFR, where the SFR is taken as the proxy for time. In this diagram, we 
show evolutionary tracks for the clouds with the various accretion rates.  
It can be seen in this figure that the evolutionary track for the cloud with 
$\dot M = 4.5 \times 10^3 \Msun$ Myr$^{-1}$ passes nearest the two 
loci of the Pipe and G14 in this diagram. However, it takes the model 
$\sim 5$ Myr to go from a Pipe-like stage to a G14-like stage. In the 
simulation, 2 Myr are enough to observe this transition. However, we
should take into account that the evolution of the simulated clump is 
representative of the substructure in both observed clouds, such as 
their clumps and filaments, and not to the clouds themselves. Then, the 
estimation of the model presented in this section is in qualitative 
agreement with the results from the simulation, and reinforcing the 
suggestion that the Pipe would evolve into a G14-like stage in the course
of some megayears if the Pipe were embedded in an environment similar 
to that of G14.

According to the model (see Fig.\ \ref{fig:zamoramodel}) the evolution 
of G14 is consistent with our prediction based on the energy budget 
evolution observed in the $\Lcal-\Sigma$ diagram. These two evolutionary 
predictions show that low-mass clouds are able to evolve to high mass 
regions in a few Myr if their external mass accretion rate is large enough.

 
 \section{Discussion}\label{s:discussion} 
 
\subsection{Implications} 
 
The evolution of the numerical clump sample studied in this work is 
consistent with the analytical calculations presented in BP+18, 
in the sense that the clumps evolve in the $\mathcal{L}-\Sigma$ diagram 
from the sub-virial region to the energy equipartition region, similarly to 
the evolution predicted by eq.\ (\ref{eq:Lcal-Sigma}).\footnote{In a recent 
numerical study, \citet{Padoan16} suggest that clumps follow the standard 
Larson linewidth–size relation, regardless of the column density. However, 
on the one hand, this study has been performed in a numerical simulation 
with supernova driving into a closed box of size 256 pc per side, which 
does not allow the hot gas to escape to high altitude above the galactic plane, 
thus raising concerns that their numerical box is over-driven, as discussed 
in \citet{Camacho16}. On the other hand, the column densities of the clumps 
considered by \citet{Padoan16} are rather low ($\Sigma \lesssim 100 \Msun$ pc$^{-2}$), 
and still coincide with the range where indeed little or no correlation between 
$\Lcal$ and $\Sigma$ is observed \citep[e.g.,] [] {Leroy15, Traficante18a}. 
It is unclear whether clumps in \citet{Padoan16} with densities 
$\Sigma \gtrsim 100 \Msun$ pc$^{-2}$, comparable to those in G14, 
will anyway exhibit the expected $\Lcal$-$\Sigma$ trend.}
However, the clumps in the present study show a slightly different behavior. 
In our case, the clumps approach the equipartition lines almost perpendicularly
to them, while the prediction is that they should approach these lines asymptotically 
(cf.\ eq.\ (\ref{eq:Lcal-Sigma}) and the left panel of Fig.\ 1 of BP+18). 
Moreover, the column density of our clumps does not increase 
monotonically over time, as in Fig.\ 1 of BP+18, but rather decreases 
again at late stages during the clumps' evolution (see top panels of Figs.
\ref{fig:simsfe} and \ref{fig:simsfr}). This difference is mainly due to the 
definition of the clumps employed in each case. In our numerical sample, 
clumps are defined as connected regions above fixed thresholds in the 
density field. This definition implies that the clumps are free to vary in size 
and mass throughout  their evolution, while in BP+18 clumps have 
constant mass by construction.  
Additionally, the prescription of sink formation above a certain critical 
density precludes the formation of regions with density above the critical 
sink-formation value. Then, as the clumps evolve, they become denser 
until sink formation becomes significant. On the other hand, the analytical 
treatment in BP+18 simply follows spherical, constant-mass clumps as 
they collapse and increase their column density. Moreover, loss of gaseous 
mass to star formation is not considered in the analytical treatment of 
BP+18. Therefore, in this idealized setting, clumps can only decrease 
their volume and increase their column densities as they collapse. 
Instead, our numerical clumps can lose mass by forming sinks, and 
when they do, their density and column density decrease. Therefore,
their evolutionary tracks in the $\mathcal{L}-\Sigma$ diagram differ from 
the analytical treatment, as also observed in the numerical core sample 
considered by BP+18, which describes paths in the $\mathcal{L}-\Sigma$ 
diagram similar to those of our own sample, characterized by a 
turnaround. 
On the other hand, the data from our numerical clumps and cores are 
also consistent with the observational data. Early in their evolution, the 
substructure in the numerical cloud exhibits a low column density, in the 
same range as that of the cores in the Pipe cloud, which is thought to be 
at an early evolutionary stage. Similarly, at later times, the substructure in 
the numerical cloud occupies the same $\Sigma$ range as the 
substructures in the G14 cloud.

Another important point to notice is that the ensemble of clumps in our 
simulation does contain a fraction of super-virial clumps at low column 
densities in the $\mathcal{L}-\Sigma$ diagram, or low masses in the 
$\alpha$-$M$ diagram, similarly to the case of observational surveys
\citep[e.g.,][] {Kauffmann13, Leroy15}. This was interpreted in 
\citet{Camacho16} and BP+18 as a consequence of those clumps being 
either in a dispersal state, or in an assembly stage by external (non-self-
gravitating, or ``inertial'') compressions or by a large-scale potential well 
which drag them gravitationally. In neither case do the inertial motions 
(by ``turbulence'' or by an external gravitational field) provide support 
for the clumps to be in a near-equilibrium state, and so, for super-virial 
clumps, there is no need for an external confining thermal pressure.
However, for clumps undergoing assembly \citep[roughly half of the clump 
population;][]{Camacho16}, the inflow may effectively be considered as 
a ``confining'' ram pressure for the densest material.  A large range of 
$\alpha$, from sub- to super-virial, is also observed in our sample of cores 
in the Pipe. Despite being in the low-$\Sigma$ range, roughly half of the 
Pipe cores show high-$\alpha$ values, which, as discussed above, can be 
an indicator of assembling or dispersing motions, rather than a confining 
pressure.

As discussed by \citet{Camacho16} and BP+18, the starting location in
the $\mathcal{L}-\Sigma$ diagram of clumps undergoing inertial assembly 
must appear as super-virial, since, by definition, the inertial assembly 
speed for these objects is larger than their self-gravitating speed, 
$\sigmag$, given by eq.\ (\ref{eq:vinfall}). Later, as the clump becomes
denser and more massive, and its self-gravitating speed increases, the 
latter eventually becomes larger than the assembly speed, and the clump 
may appear to be either in equipartition or sub-virial when it becomes
dominated by self-gravity. In particular, it will appear sub-virial if $\sigmag$ 
is still smaller than the equipartition value 
$\sigmaeq \equiv \sqrt{\eta GM/R}$, because $R$ is still not sufficiently 
smaller than $R_0$ (upper solid lines in the left panel of Fig.\ 1 of BP+18).
On the other hand, clumps can exhibit sub-virial Larson ratio or virial 
parameter during the early stages of their contraction if their initial internal 
turbulent velocity is low, and their infall speed has not yet reached the free-fall 
value \citep{Ballesteros18}. In addition, \citet{Traficante18b} have suggested 
that a sub-virial appearance can also occur in observations as a consequence 
of a mismatch between the regions from which the mass and the velocity 
dispersion are derived, if these are selected using different tracers. However, 
this effect cannot be at play for the data from our simulation, in which the clump 
properties are measured directly from the numerical data, and therefore their 
sub-virial nature has to be a real physical property.

The fact that the numerical clumps investigated in this work initiate
their trajectories in the sub-virial region of the $\mathcal{L}-\Sigma$ 
and $\alpha$-$M$ diagrams indicates that they are already dominated
by self-gravity, yet they have not had time to attain the equipartition speed, 
since equipartition is only attained at later times in their evolution, as
shown by Figs.\ \ref{fig:khobs} and \ref{fig:obs}. In the observational 
sample, this increasing-$\alpha$ evolutionary trend is shown by the 
absence of sub-virial objects in the G14 cloud.

It is important to note also that, were we to follow the evolution of our 
numerical clumps to even more advanced stages, they would move 
into the super-virial region, since they would have lost a large amount 
of mass to sink formation, therefore losing gravitational energy from 
the gas mass, but the velocity dispersion would remain roughly the 
same. Therefore, they would appear to be super-virial. 
In \citet{Camacho16} we showed that some apparently super-virial clumps
could be made to appear in equipartition again when the mass in sinks
was included in the computation of the kinetic to gravitational energy
balance. Finally, this effect would also occur if feedback were included, 
because in this case mass would be lost from the clumps due to the 
feedback, also reducing their gravitational content. Thus, apparently
super-virial clumps are likely to occur once a significant amount of their
gas mass has been converted to stars and/or feedback has expelled a
significant amount of mass from them. All of this, aside from geometrical 
effects such as those discussed by BP+18.\footnote{In BP+18 it was 
shown that $\alpha$ computed with the standard definition given by the 
second equality of eq.\ (\ref{eq:alpha}) can often overestimate the true 
value of the energy ratio, because it assumes that the gravitational energy
is given by the expression for uniform density sphere, which is smaller than 
the true gravitational energy of a centrally concentrated object.}

Finally, as mentioned above, we have been able to interpret the  physical 
properties and star formation activity of two different star forming regions 
as a consequence of evolutionary effects that transform one into the other 
in a simulation including only decaying turbulence and self-gravity. 

\subsection{Caveats}
\label{sec:caveats}

The simulation presented in this work clearly lacks important physical 
ingredients, most notably, stellar feedback and magnetic 
fields. The omission of stellar feedback, however, is probably not a major 
concern, as discussed in detail in Appendix \ref{app:feedback}. Indeed, 
our region of study  (``Region A") has been chosen to be far from any regions 
harboring massive stars that could produce supernova remnants capable 
of disturbing the region at the time studied, and at a time in its evolution 
at which no massive stars have formed yet that could affect it by photoionizing 
radiation. Feedback from low-mass stars is well known to possibly affect the 
structure only on $\sim 1$ pc scales \citep[e.g.,] [] {Bally16}, but not at the 
cloud scales, $\sim 10$ pc.

For the same reason, we avoid using artificial turbulence driving (generated, for example, in Fourier space). On the one hand, Appendix B shows that our region is unlikely to be affected by feedback during the time interval in which we study it. On the other hand, even if feedback had to be included, Fourier driving is rather inadequate as a substitute, since it is applied everywhere in space and nearly continuously in time, rather than intermittently in both time and space, as real stellar feedback is.

Concerning the magnetic field, it is now well established that most clouds 
and dense cores are magnetically supercritical \citep[e.g.,] [] {Crutcher12}, 
and thus evolve under the domination of self-gravity, although the presence 
of a weak field could still slow down the gravitational contraction. 
We plan to repeat our study in the presence of a magnetic field in a future paper.

On the other hand, in spite of the above limitations, our simulation considers 
features that are usually neglected in numerical studies of molecular clumps. 
Most importantly, it follows the self-consistent evolution of the clump since the 
formation of the cloud complex it belongs to, thus allowing for the growth and 
continued accretion onto the clump, which regulate its physical properties and 
star formation activity. This is the reason why our study focuses on a time more 
than 20 Myr after the start of the simulation, since it has been previously 
necessary for the parent cloud complex to form and grow from the warm diffuse 
gas, and then engage in global gravitational contraction, which is the essence 
of the GHC scenario. The clump itself follows a similar path at its own scale. 
This self-consistent evolution cannot be followed by simulations of isolated, 
isothermal regions over short time spans. We consider that this setup is ideally 
suited for studying the early evolution of clumps undergoing GHC.

\section{Summary}\label{s:summary}

In this work we  have presented a study of the energy budget of a star-forming 
clump in a numerical simulation in which many cloud complexes form by global 
hierarchical collapse, and compared the numerical results to data from two real 
clouds (Pipe and G14). The clumps studied in this work, both from 
the simulation and from the observational data, lie close to the energy 
equipartition relation, shown in the $\Lcal-\Sigma$ diagram. This trend 
has been approximately found in several previous observational and 
numerical studies \citep{Keto86, Heyer09, Dobbs11,Leroy15,Traficante15, 
Camacho16, Ballesteros18}. However, deviations from it are also 
systematically observed. In this study, we suggest that these deviations 
are characteristic of the clump evolutionary state, and that this evolution 
determines also the star formation activity of the clumps.

The main result from the present study is that our numerical clump evolves 
from a Pipe-like state to a G14 one in roughly 2 Myr. The Pipe-like stage 
is characterized by a lower mass, velocity dispersion, mean and peak 
densities, and star formation activity, while maintaining comparable 
dimensions. The increase in physical parameters is due to accretion of 
material external to the $\nth = 10^3 \pcc$ boundary we used to define 
this clump, as indicated by the fact that the time delay for the appearance 
of the first stars after the time at which we first observe the clump is within 
$\sim 10\%$ of the free-fall time for the clump at the starting time. In 
addition, a comparison with the analytical model by \citet{Zamora14}, 
which describes the evolution of the physical parameters and star-forming 
activity of a gravitationally contracting cloud, using initial parameters
appropriate for the Pipe cloud, shows that a G14-like cloud (in mass and 
SFR) appears roughly a few Myr later. These results strongly suggest that 
clouds evolve from low-mass, low-density, and low-SFR states to states of 
higher masses, densities and SFRs over the course of a few megayears if 
their external mass accretion rate is large enough.

This evolution of the clouds implies an evolution of the energy budget
of their substructures, so that samples of cores of younger clouds
appear displaced toward more sub-virial states in the
$\Lcal$--$\Sigma$ and $\alpha$-$M$ diagrams, and, as the clouds age,
their structures are displaced in these diagrams towards higher-column
densities and closer to equipartition, due to the mechanism described
in BP+18 and \citet{VS19}, which predicts that the  Larson ratio and the 
virial parameter of cores that decouple from the global flow and begin to 
contract at a finite radius $R_0$ evolve during the core's contraction. The 
details of the evolution depend on the initial ratio of the inertial external 
compressions to the self-gravity-driven motions approaching the 
equipartition values as collapse proceeds and the inertial motions 
become subdominant  \citep[see Fig. 2 of ][]{VS19}.
Nevertheless, at late stages, coeval samples of clumps and cores exhibit 
the regularly-observed feature that more massive objects have lower 
values of $\alpha$. 
\citet{VS19} have suggested that the lower $\alpha$ of more massive 
objects may arise as a consequence of them having lower average 
densities, and therefore longer free-fall times, causing them to evolve 
more slowly in their approach to equipartition.

Our results lead us to  suggest that massive dense cores that appear
sub-virial and quiescent \citep[i.e., prestellar or very weakly-star-forming 
cores; e.g.,] [] {Kauffmann13, Liu15, Ohashi16, Sanhueza17, Contreras18} 
will evolve toward equipartition as they develop stronger star-formation 
activity. Also, since we found that, at early stages, the densest parts of the 
clumps appear more sub-virial than their envelopes, we predict that 
observations of the parent clumps of those sub-virial cores should reveal 
that the parent structures are closer to equipartition.\\

\section*{Acknowledgments}
The authors thank the anonymous referee for the useful comments that helped to improve this work.
V.C. acknowledges support from CONACyT grant 406297.
V.C. and EVS are thankful to the project CB-2015-255295 supported by CONACyT. 
A.P. acknowledges financial support from UNAM-PAPIIT IN113119 grant, M\'exico.
G.B. is supported by the MINECO (Spain) AYA2017-84390-C2 grant.
M.Z.A. acknowledges support from CONACyT grant number A1-S-54450 to Abraham Luna Castellanos (INAOE).

\software{
Astrodendro (\url{https://github.com/dendrograms/astrodendro}),
Astropy \citep[\url{http://www.astropy.org},][]{Astropy13, Astropy18}
CASA \citep{McMullin07}, 
GADGET2 \citep{Springel05},
GILDAS \citep[\url{www.iram.fr/IRAMFR/GILDAS},][GILDAS team 2013]{Pety05}.
}

\appendix

\section{Testing Dendrogram parameters } 
\label{app:dendro}

As described in Sec.\ \ref{subsec:g14procedure}, for selecting the structures in G14 we used the \dendro\ algorithm. However, the objects defined by this algorithm depend heavily on the values of the parameters chosen \citep[e.g., ][]{Goodman09, Burkhart13}. Specifically, two crucial parameters are {\tt min\_value} and {\tt min\_delta}. These respectively define the minimum intensity value above which structures are searched and the ``tolerance'' allowed for fluctuations above a certain level to identify a structure as an independent one. Thus, {\tt min\_value} is equivalent to the density threshold we use to define objects in the simulation and {\it defines the class of object being considered.} That is, successively higher values of \ttmv\ allow defining clouds, filaments, clumps or cores. On the other hand, \ttmd\ determines how strong a neighboring peak needs to be to be classified as a separate structure. In particular, this determines whether moderate-amplitude fluctuations around a large structure are interpreted as separate structures or incorporated into it. That is, this parameter determines whether a given structure ``branches out'' into smaller ones or not. In turn, this causes the isocontours of a given large structure to increase in size when \ttmd\ is increased, since the structure will then incorporate additional minor structures around it. This non-uniqueness of the isocontours that can be defined by the \dendro\ algorithm is a direct consequence of the attempt to define isolated objects amid a continuous medium. Therefore, the question arises\footnote{We thank the referee for pointing this out.} as to what is the uncertainty in the values of the virial parameter and the Larson ratio introduced by the ambiguity in the definition of the structures.

To estimate the uncertainty in our results introduced by the choice of parameters, in this Appendix we consider an alternate choice of the \ttmd\ parameter, which affects the isocontours of the structures we identified in G14. We do not vary \ttmv, since it only determines the class of object, and is thus as arbitrary as our choice of density threshold for the simulation data.

\begin{figure}
\begin{center}
\subfloat[]{\includegraphics[width = 3.2in]{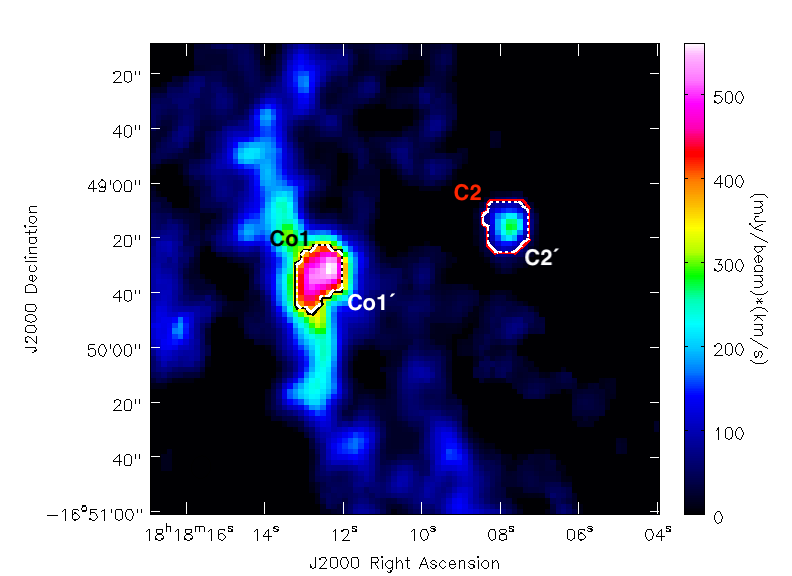}}\label{fig:16a} 
\subfloat[]{\includegraphics[width = 3.2in]{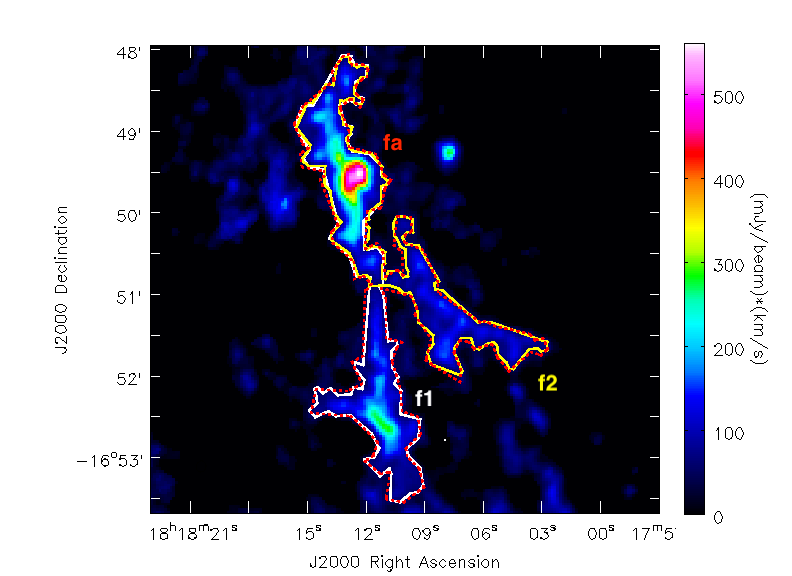}}\label{fig:16b}\\
\subfloat[]{\includegraphics[width = 3.2in]{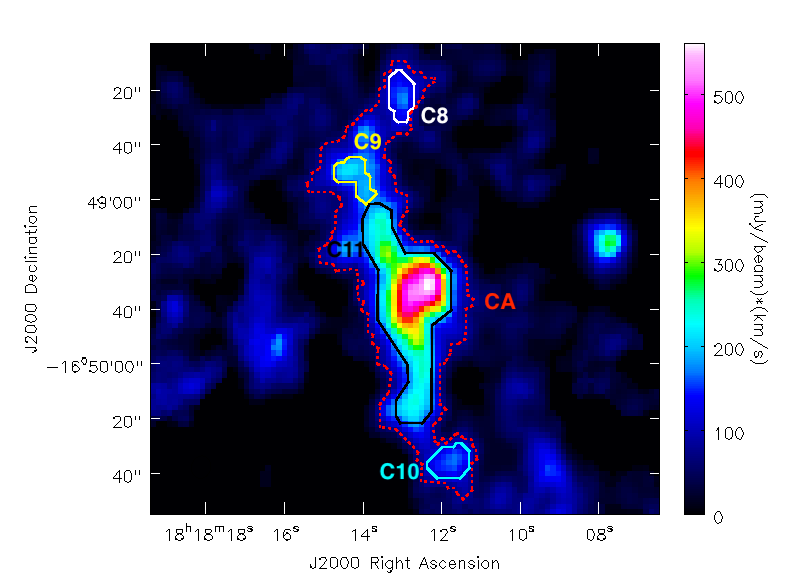}}\label{fig:16c}
\subfloat[]{\includegraphics[width = 3.2in]{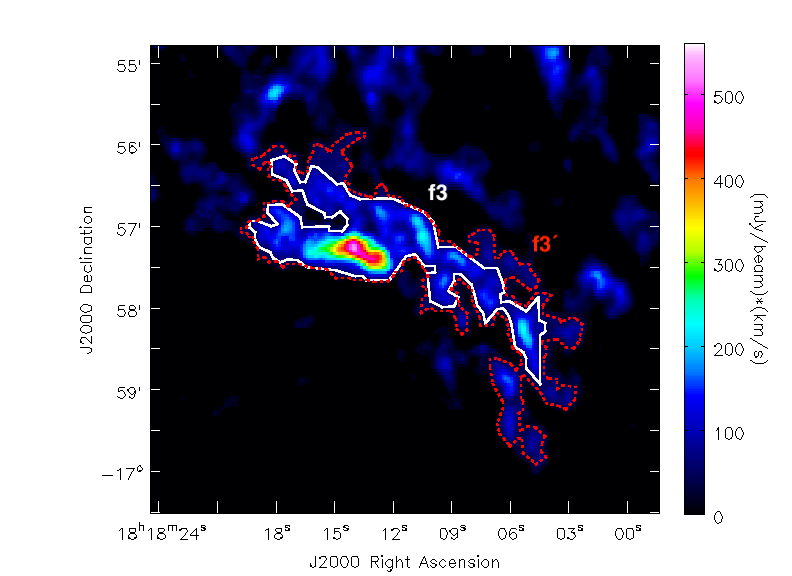}}\label{fig:16d}\\
\subfloat[]{\includegraphics[width = 3.2in]{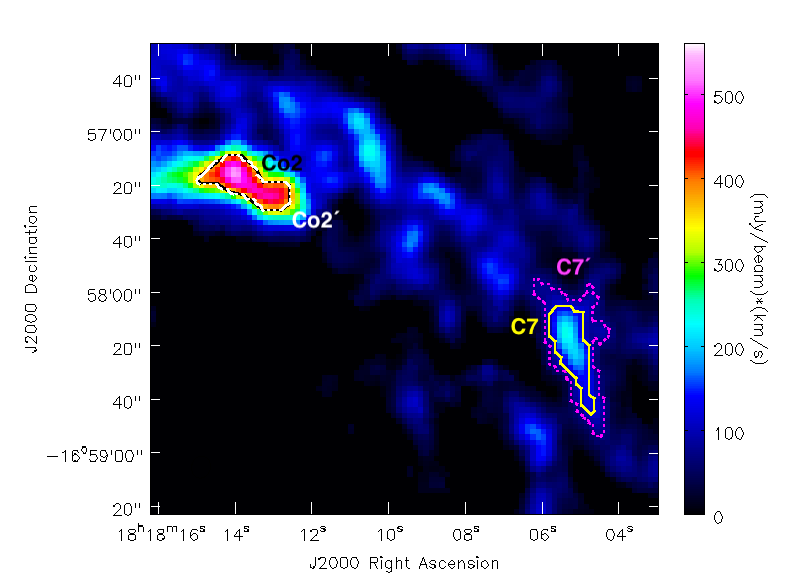}}\label{fig:16e}
\subfloat[]{\includegraphics[width = 3.2in]{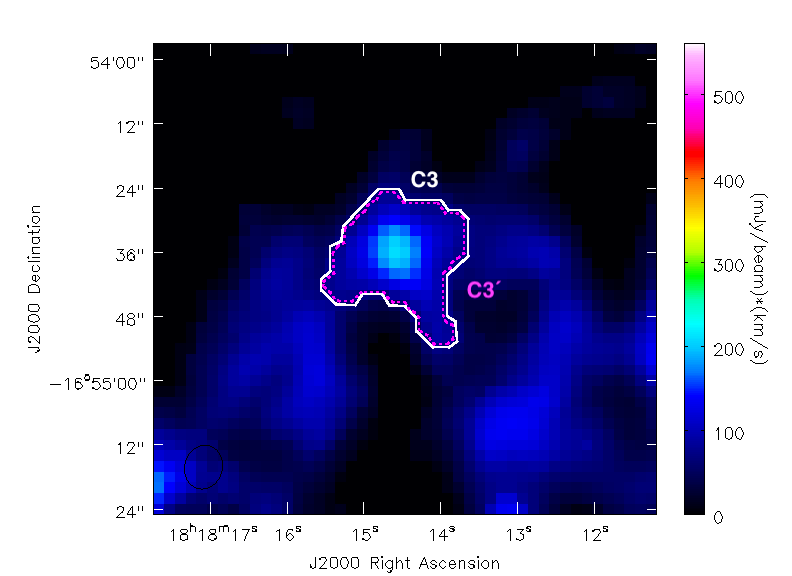}}\label{fig:16f}
\end{center}
\caption{Comparison between the original G14 sample (solid lines) and the structures obtained with \texttt{min\_del} = 5$\sigma$ (dotted lines). }
\label{fig:testg14sample}
\end{figure}

In Sec. \ref{subsec:g14procedure} we discussed the selection of the \texttt{min\_value} and \texttt{min\_delta} values for each of the defined kind of objects: cores, clumps, and filaments, which are (\texttt{n}$_{\rm v} $, \texttt{n}$_{\rm d}$)= (15,3), (5,1), and (1,1) respectively. We will refer to this set of structures as Sample I. For the test, we now choose a fixed value \texttt{n}$_{\rm d}$=5 for each class in order to reduce the branching. 
Given that \texttt{min\_value} remains the same as in Sample I, the increase of \texttt{min\_delta} produces a new set of objects, which we refer to as Sample II. 
From it, we select a subgroup of objects related to those of Sample I (see Fig. \ref{fig:testg14sample}), to investigate  how their properties change.

To explore the differences between the same kinds of objects, we show in Fig.\ \ref{fig:testg14sample} the contours defined by \dendro\ in the original G14 sample (Sample I; solid lines) and those obtained with the new value of \ttmd\ (Sample II; dotted lines). The labels are the same as in Fig.\ \ref{fig:g14sample} for the objects in Sample I, while those from Sample II are labeled with the same name with an apostrophe for those objects having a counterpart in Sample I, and with a letter for those that are different. It can be noticed in Fig.\ \ref{fig:testg14sample} that objects in panels (a) and (f), and object ``Co2" in panel (e)  have nearly the same contours in both samples, which indicates that these compact objects do not depend significantly on the value of \texttt{min\_delta}. This is because the density profile in these objects is very steep, and no new material is incorporated into them by increasing \ttmd.
On the contrary, the rest of the objects in Sample II show significant variations in their contours. In particular, note that many of the peripheral structures from Sample I are incorporated into objects from Sample II. For example, filament ``fa"  from Sample II contains  filaments ``f1" and ``f2" from Sample I, while clump ``CA" in Sample II contains clumps ``C8-- to ``C11" from Sample I. Also, the counterparts of ``f3" and ``C7", in the new sample include small structures discarded in Sample I. 

From the images in Fig.\ \ref{fig:testg14sample}, it is clear that the increment in \texttt{min\_del} causes the small structure to be absorbed in the new sample, see for example, ``f3" and ``C7". Another effect when changing this parameter is the generation of a new hierarchy of objects, for example ``CA", which has been defined with the \texttt{min\_value} for clumps in Sample I, but with the new \texttt{min\_del}, it should be defined as a filament. Thus, it is expected that their corresponding physical properties, in particular $\Lcal$ and $\alpha$, must change.

\begin{figure}[]
\begin{center}
\subfloat {\includegraphics[width = 3.5in]{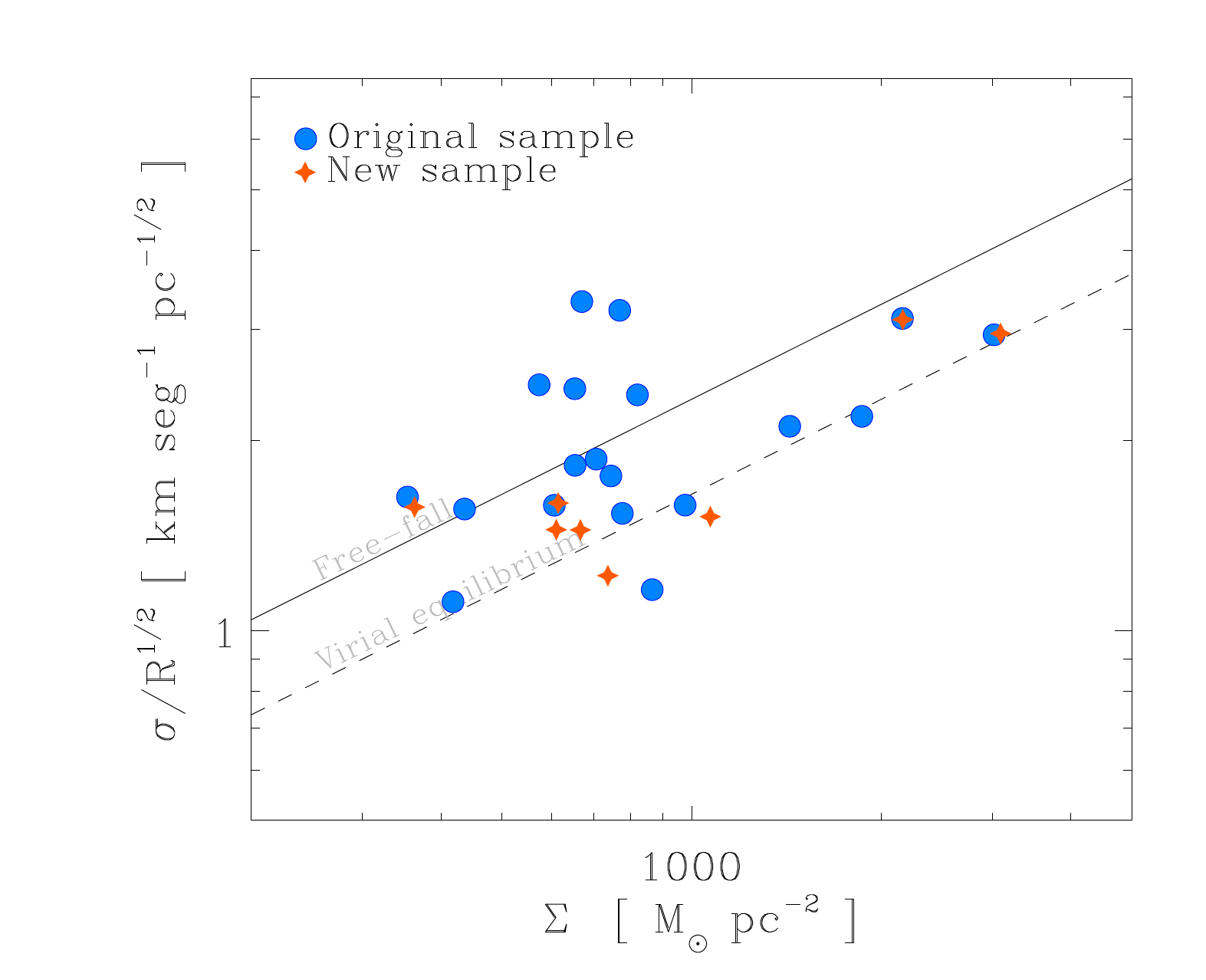}} 
\subfloat {\includegraphics[width = 3.5in]{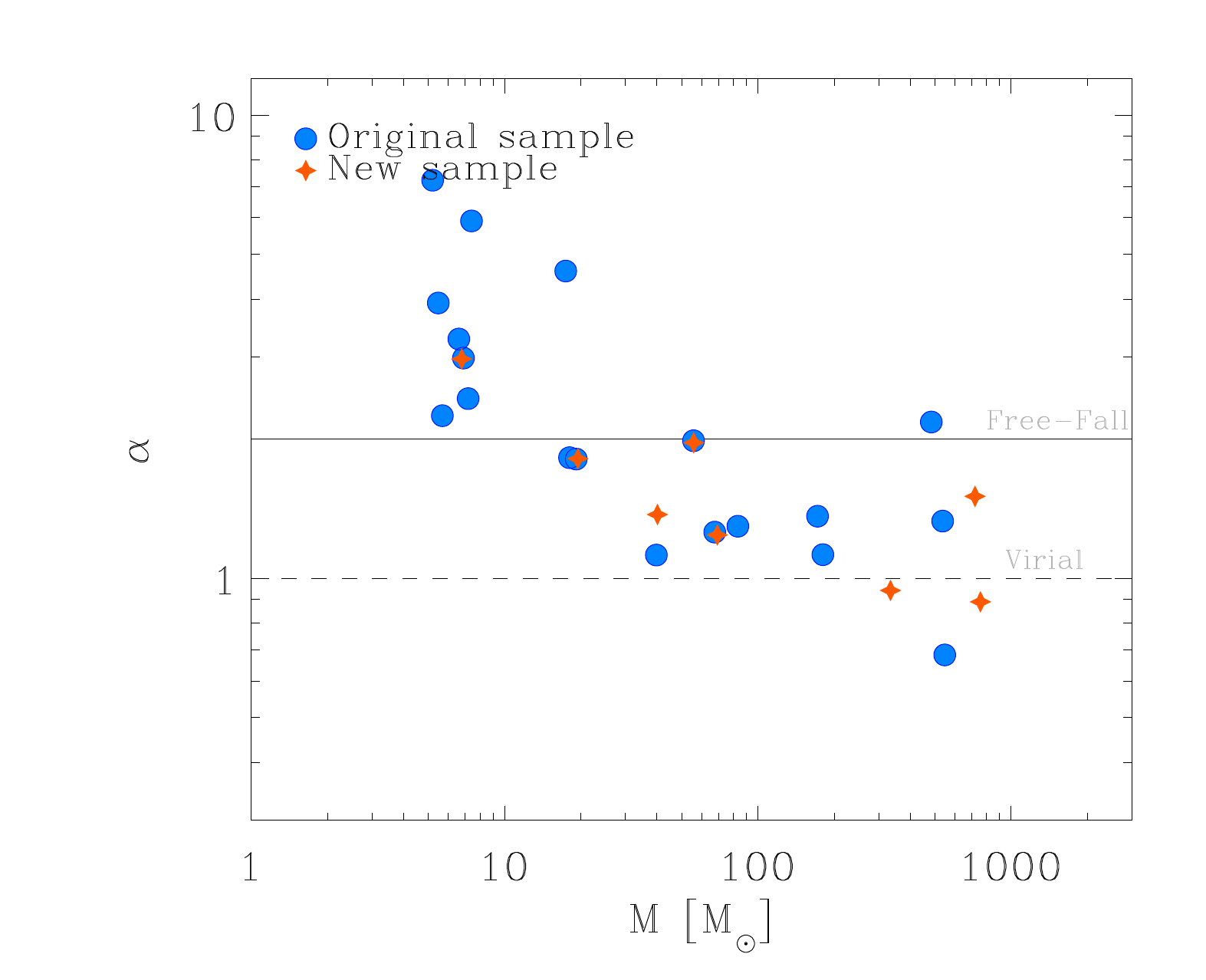}}
\caption{$\Lcal-\Sigma$ plot (left panel)  and $\alpha$ plot (right panel) for the original G14 sample (Sample I, blue circles) and the new structures obtained with \texttt{min\_delta}= 5$\sigma$  (Sample II, orange stars). Both samples are located in the same locus, suggesting that there is no significant qualitative variation when changing the \dendro\ input parameters. }
\label{fig:testnewobjects}
\end{center}
\end{figure}

\begin{figure}
\subfloat{\includegraphics[width = 3in]{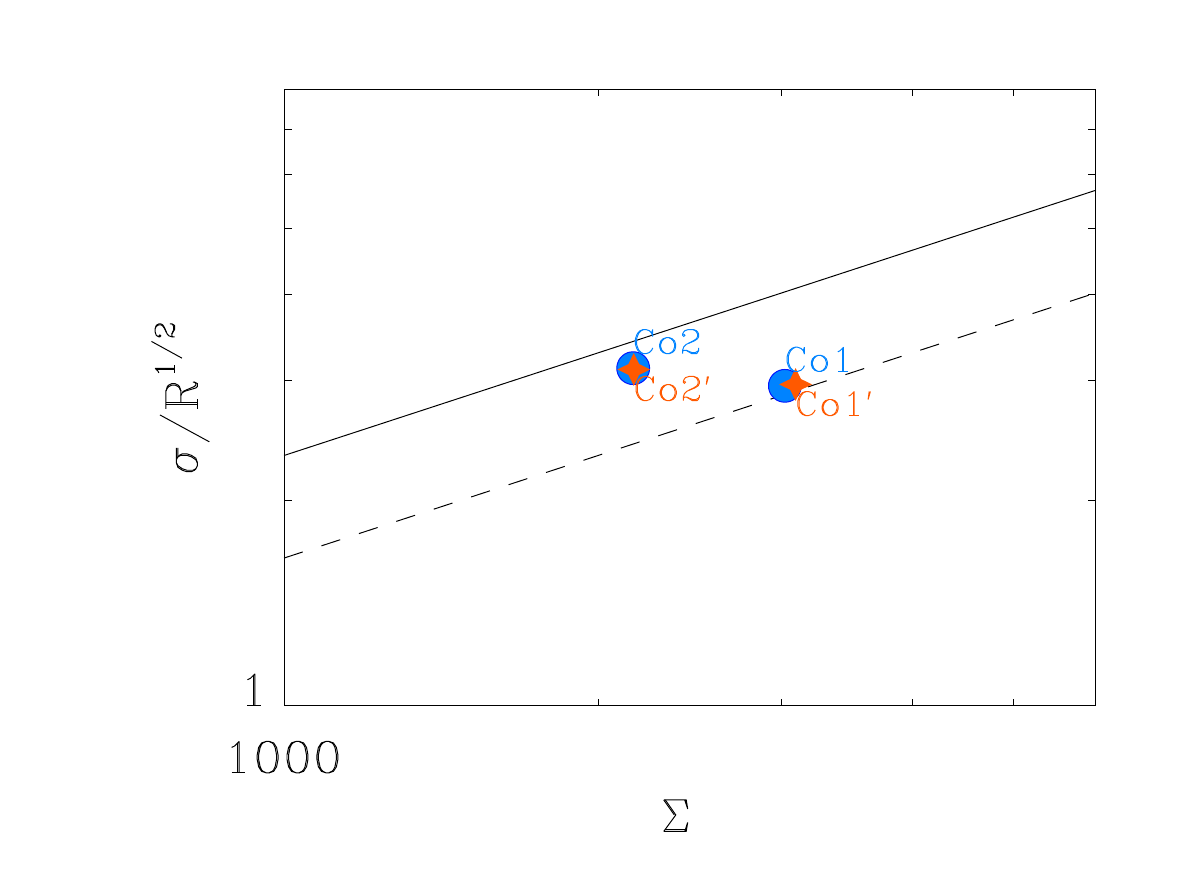}} 
\subfloat{\includegraphics[width = 3in]{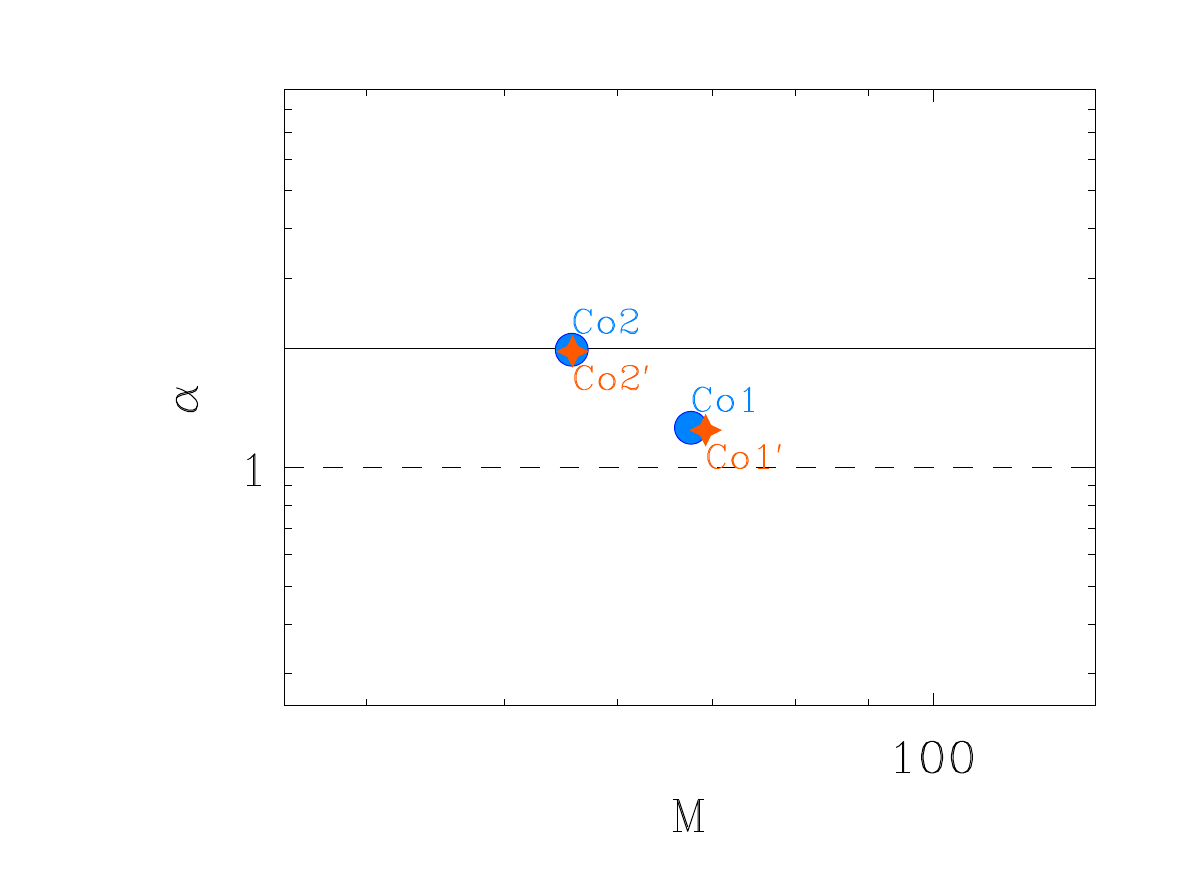}}\\
\subfloat{\includegraphics[width = 3in]{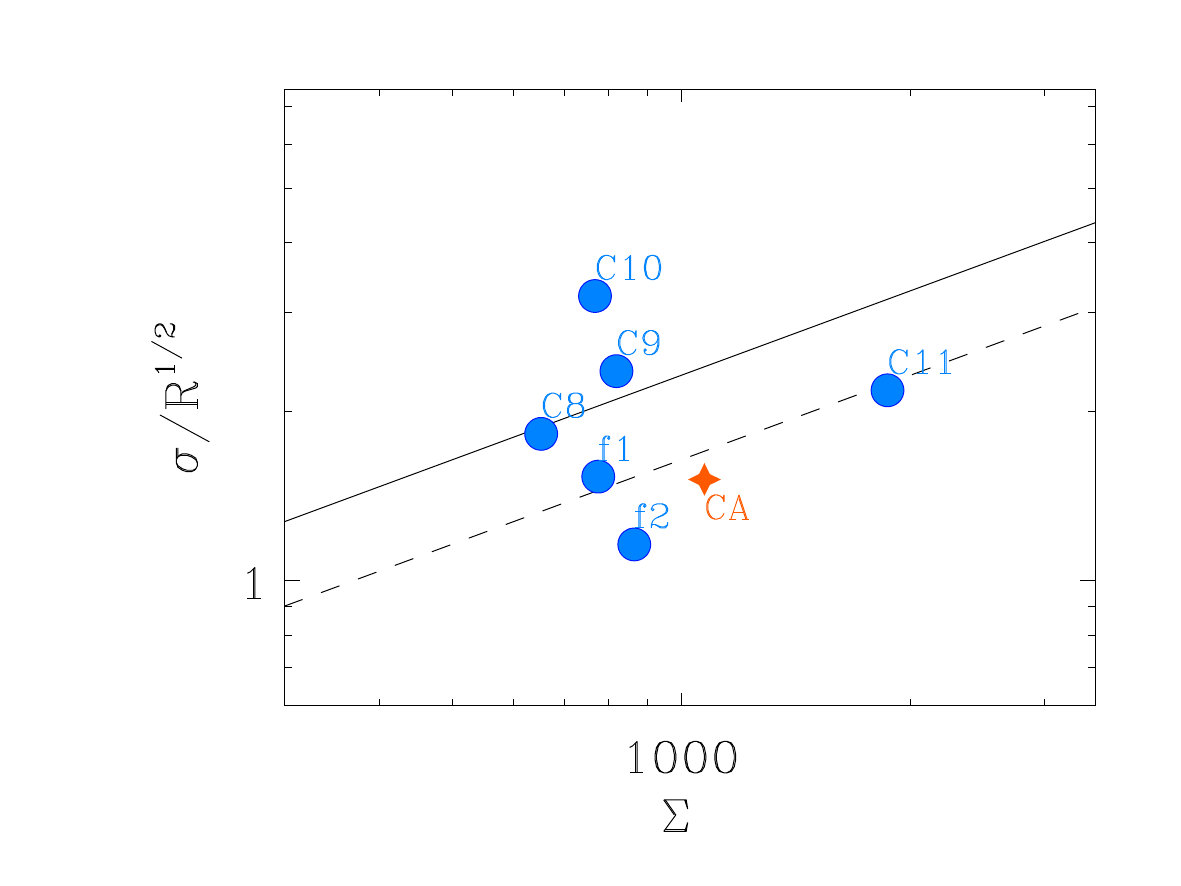}}
\subfloat{\includegraphics[width = 3in]{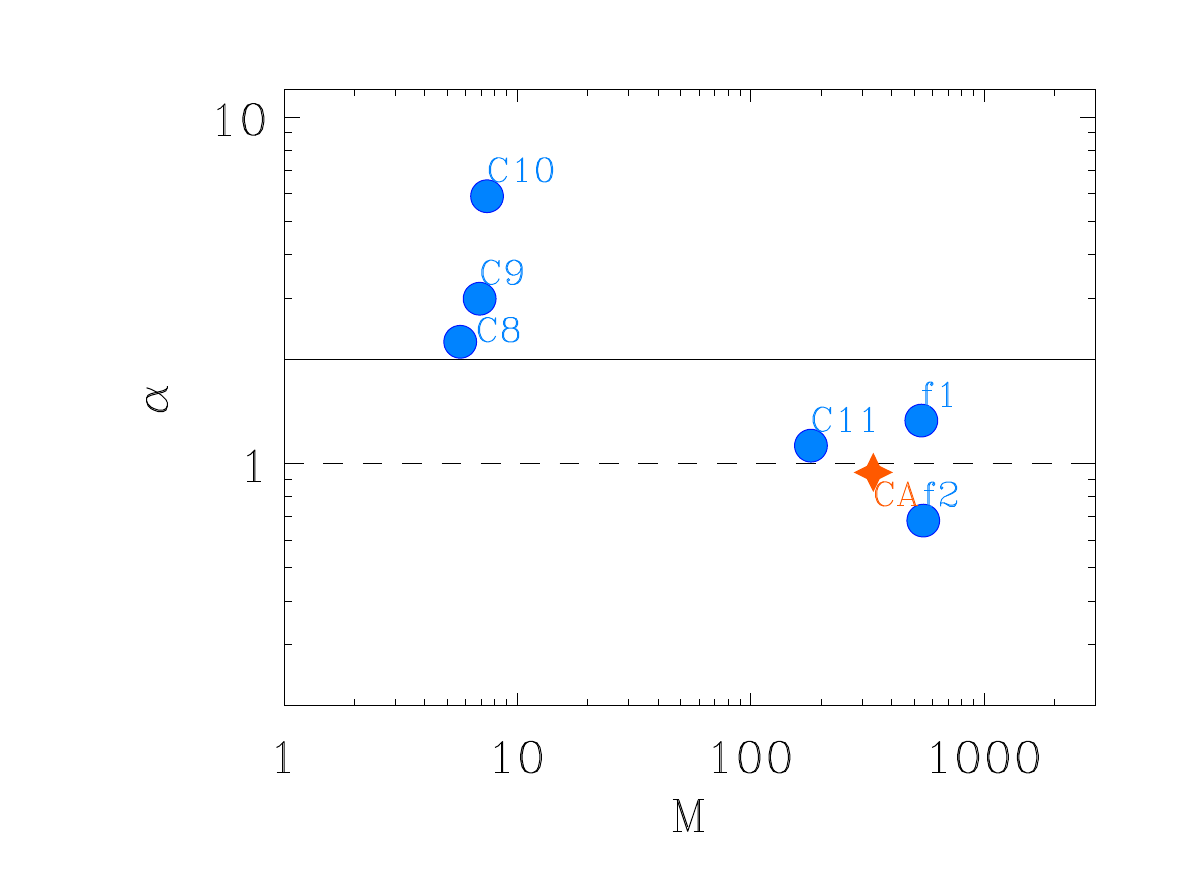}}\\
\subfloat{\includegraphics[width = 3in]{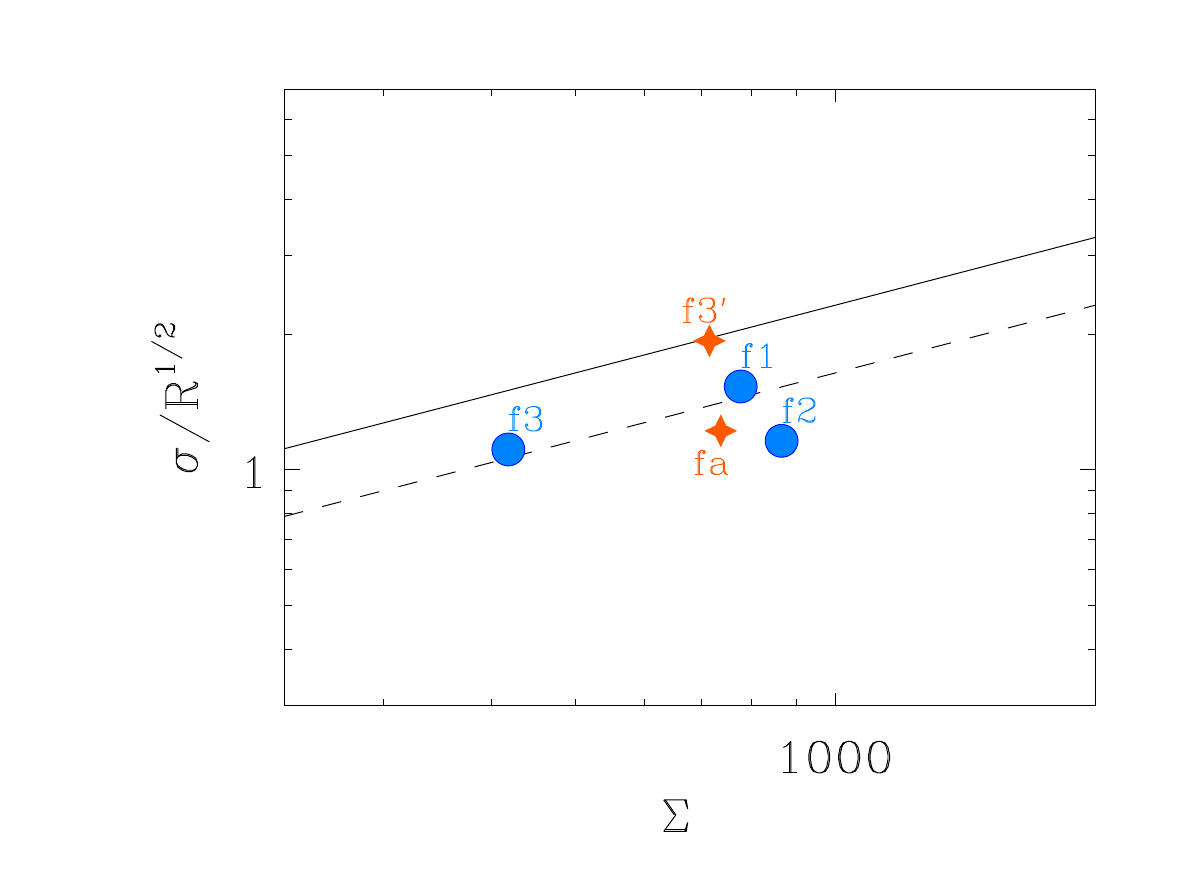}} 
\subfloat{\includegraphics[width = 3in]{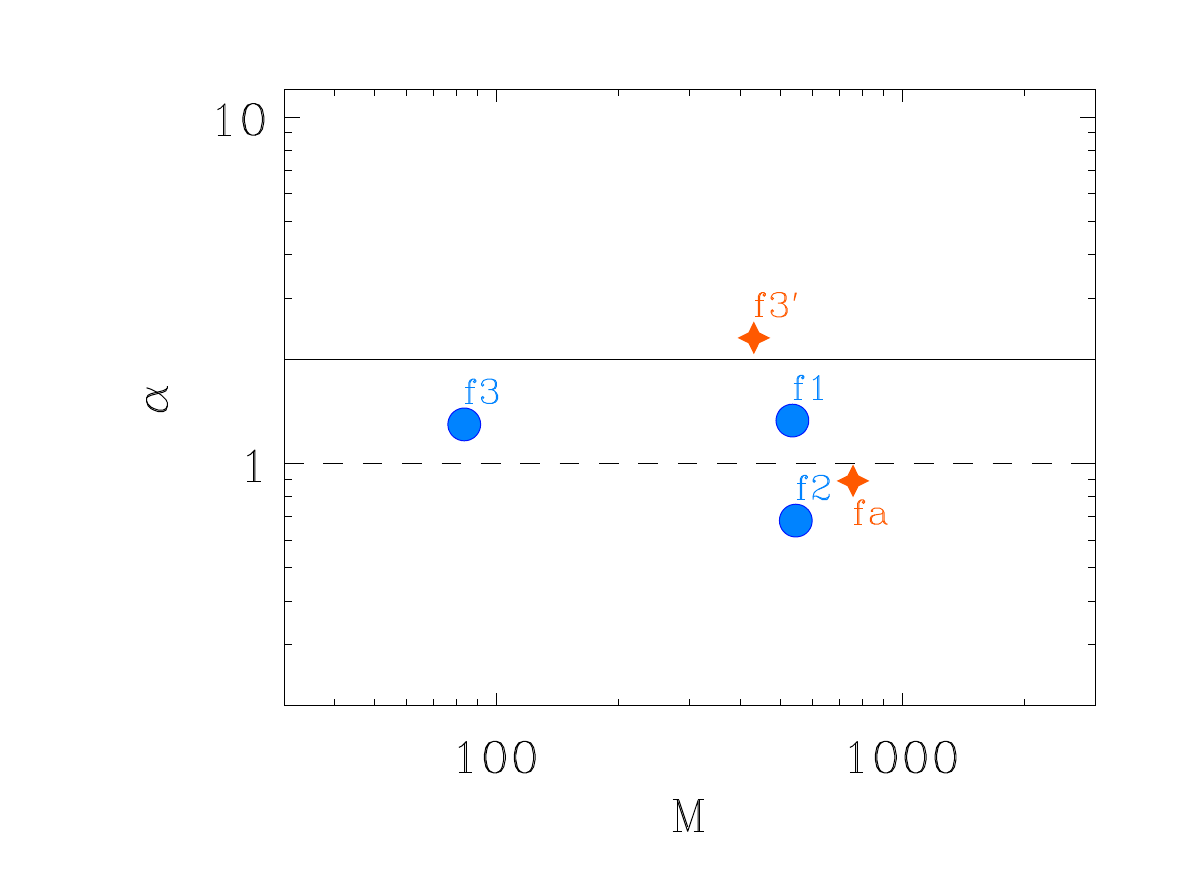}}\\
\subfloat{\includegraphics[width = 3in]{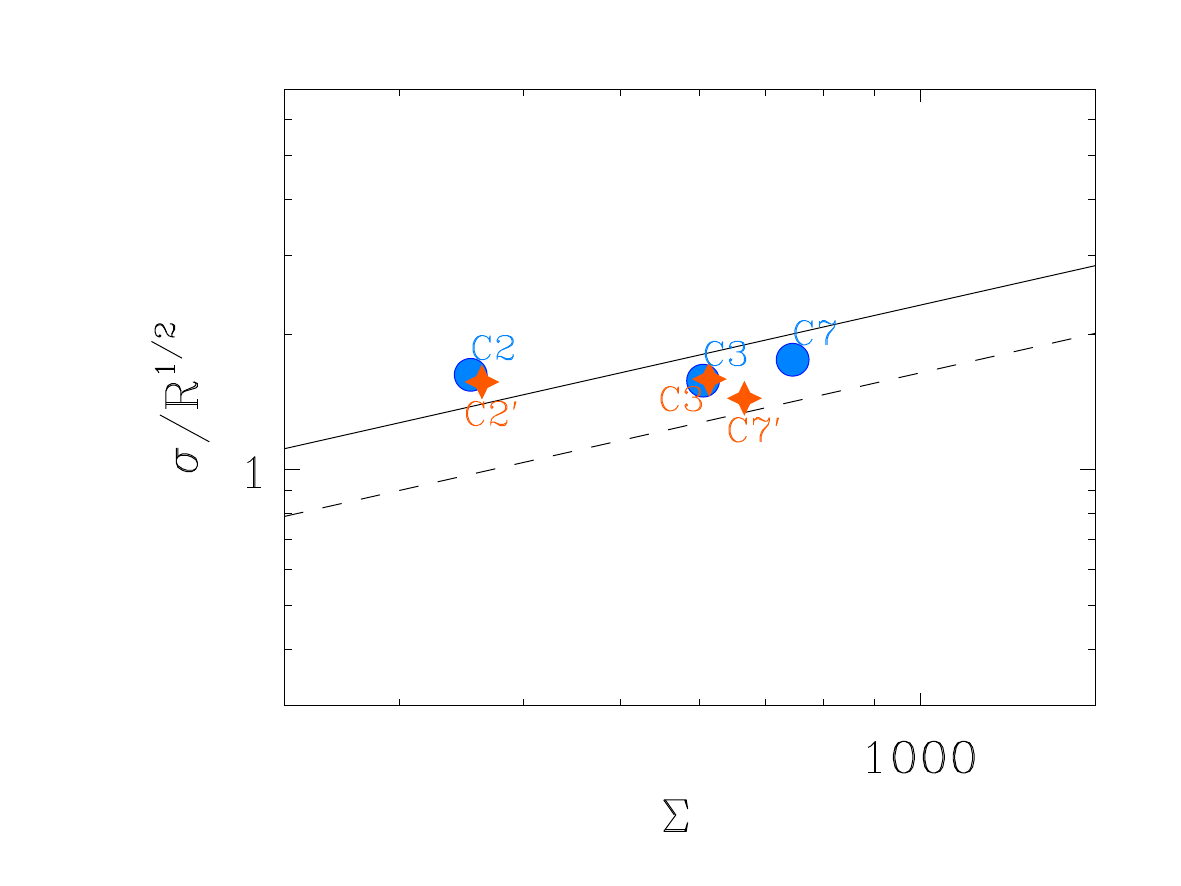}}
\subfloat{\includegraphics[width = 3in]{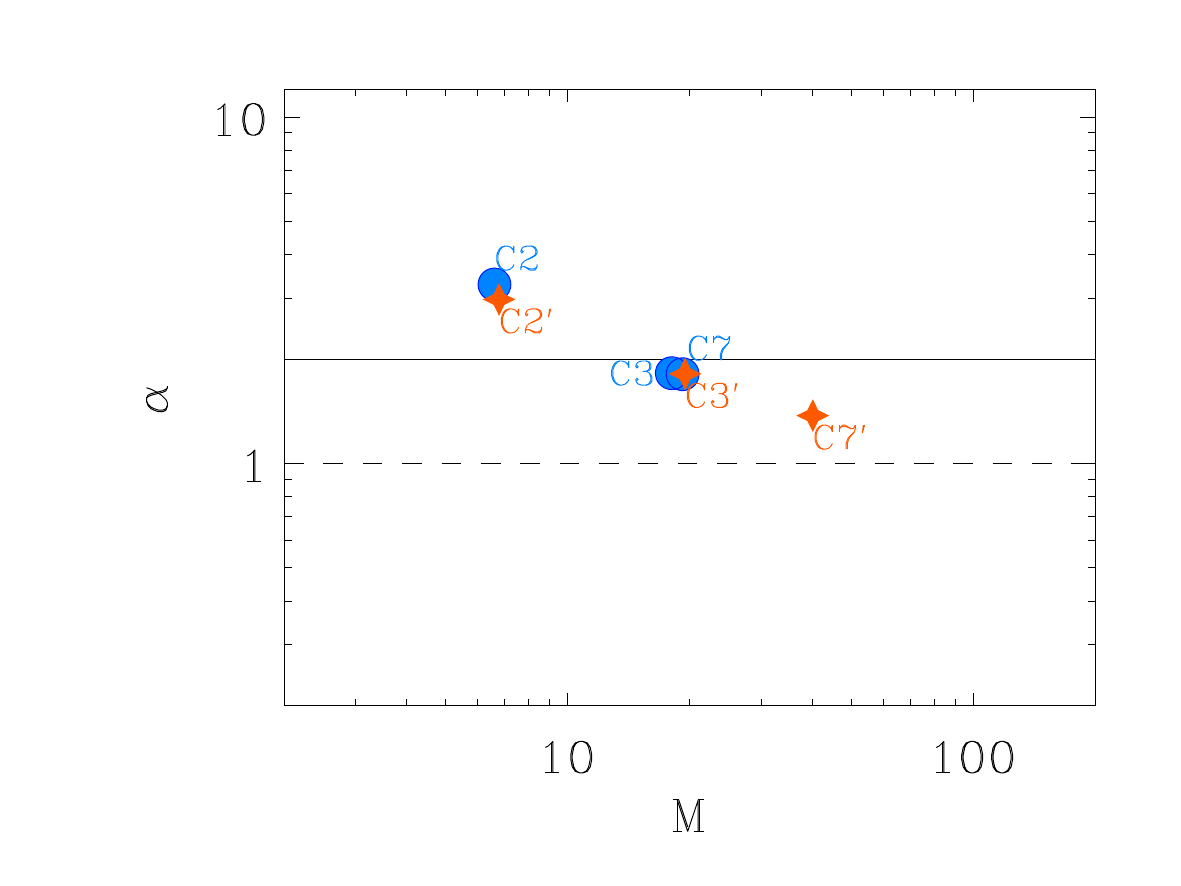}}
\caption{$\alpha$ comparison between the original G14 objects (blue circles) and the new sample (orange stars). The units are the same as s in Fig. \ref{fig:testnewobjects} and the lines represent the free fall (continue) and the virial equilibrium (dashed) conditions.}
\label{fig:testkhalpha}
\end{figure}

To investigate the effect of a different defining criterion for the objects on the values of the Larson ratio and the virial parameter, Fig.\ \ref{fig:testnewobjects} shows the $\Lcal-\Sigma$ (left panel)  and $\alpha$ (right panel) plots for the clumps from Sample I (blue circles) and the new objects from Sample II (orange stars).
It is readily seen from this figure that the objects from Sample II, in both the $\Lcal$-$\Sigma$ and $\alpha$-$M$ plots, occupy the same locus as those from Sample I.
In addition, Fig.\ \ref{fig:testkhalpha} shows the $\Lcal-\Sigma$ plot (left panels)  and $\alpha$ plot (right panels) for selected individual objects in Sample I (blue circles) and their counterparts in Sample II (orange stars). Note that, as described above, some of the objects in Sample II incorporate several objects from Sample I, so the correspondence is not necessarily one-to-one. In particular, it is seen that a new object in Sample II sometimes comprises several smaller objects from Sample I, and in this case, it is more resemblant of a larger-scale class of objects. This is the case, for example of clump CA from Sample II, which engulfs clumps C8-C11 from Sample I (see the panels in the second row of Fig.\ \ref{fig:testkhalpha}), and is thus more similar to filaments f1 and f2 in Sample I. But it is seen in this case that clump CA is located nearer to the loci of filaments f1 and f2 than to the loci of the smaller clumps C8-C11. Therefore, when the effect of the change in \ttmd\ changes the category of the resulting object (from clump to filament, for example), the values of $\Lcal$ and $\alpha$ for the new object are consistent with the values for objects of the same category in the old sample. 

We thus conclude that, rather than introducing an ``uncertainty'' in the values of $\Lcal$ and $\alpha$, varying the \ttmd\ parameter simply changes the object defined for study, with the consequent change in its physical properties. This is fundamentally a consequence of attempting to define discrete objects within a continuous medium, so that when the defining criterion is changed, the resulting object defined also changes essentially. However, because the original sample contained objects of several classes (i.e., defined with different values of the \ttmv\ parameter), the changes in the object introduced by changing \ttmd\ can at most change its classification, but it continues to be located within the locus of the full sample of objects originally defined in Sample I. Thus, our results are insensitive to variations in the parameters of the \dendro\ algorithm.

\section{Comparison with the ammonia data}  
\label{App:ammoniadata}

In this work we report the results for the G14 sample from the data 
obtained directly from the H$_2$ column density map \citep{Lin17}. 
However, the data of the $\ammonia$ (1,1) and (2,2) lines 
\citep{Busquet13} allow as well the estimation of the column density, 
N(H$_2$), by applying the radiative transfer equations (see below). In what 
follows we show the comparison of N(H$_2$) and the mass derived with 
the ammonia data and the H$_2$ column density map of \citet{Lin17} .

\begin{figure}[h!]
\centering
\includegraphics[width = 12cm]{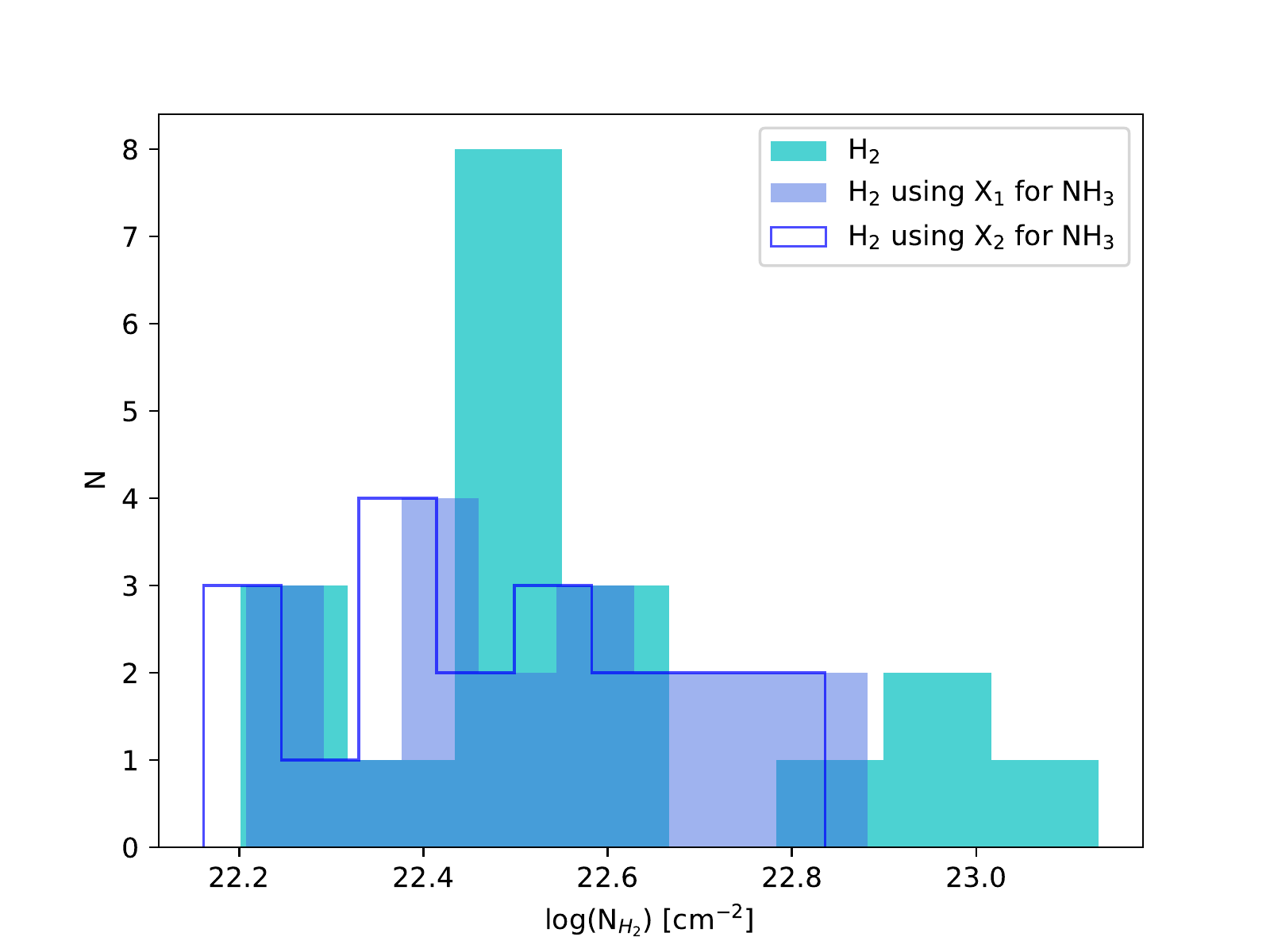}
\caption{ H$_2$ column density derived from the H$_2$ map \citep{Lin17} 
and from the NH$_3$ maps \citep{Busquet13} using two different abundances, 
X$_1=4.2\times10^{-8}$ \citep{SanchezM13} and X$_2=4.7\times10^{-8}$ 
(this work).}
\label{fig:histoN}
\end{figure}

For the computation of the physical parameters, we extracted the 
spectrum from the contours defined with the \dendro\ using the CASA 
software for both the (1,1) and (2,2) lines. We analyze the spectra with 
the CLASS ammonia method in 
GILDAS\footnote{ http://www.iram.fr/IRAMFR/GILDAS}.  
This method takes into account the hyperfine structure of ammonia. 
It computes the ammonia line profiles assuming a Gaussian velocity 
distribution and equal excitation temperatures. Then, we follow the 
procedure in \citet{Busquet09} in order to derive the column density. 
Table \ref{tab:columndensity} shows the data for the NH$_3$ column 
density derived following \citet{Busquet09} and the column density 
measured directly from the H$_2$ map \citep{Lin17}.

\begin{table}[h]
\centering
\begin{tabular}{crrr}
\hline
\multicolumn{1}{c}{ID} & \multicolumn{1}{c}{\begin{tabular}[c]{@{}c@{}}R\\ (pc)\end{tabular}} &
\multicolumn{1}{c}{\begin{tabular}[c]{@{}c@{}}N(NH$_3$)\\ $10^{14}$cm$^{-2}$\end{tabular}} & 
\multicolumn{1}{c}{\begin{tabular}[c]{@{}c@{}}N(H$_2$)\\ $10^{22}$cm$^{-2}$ \end{tabular}} \\
\hline
C1    &  0.079  & 	  7.26  & 	1.95  \\
C2	 &  0.084  & 	14.49  & 	1.58  \\
C3	 &  0.109  & 	18.59  & 	2.71  \\
C4	 &  0.107  & 	31.99  & 	2.57  \\
C5 	 &  0.054  & 	11.48  & 	3.00 \\  
C6 	 &  0.056  & 	30.56  & 	2.93  \\
C7	 &  0.096  & 	25.32  & 	3.34  \\
C8	 &  0.057  & 	 9.08   & 	2.93  \\
C9	 &  0.056  & 	10.52  & 	3.68  \\
C10	 &  0.060  & 	16.69  & 	3.45 \\
C11	 &  0.191  &  14.80  & 	   8.36  \\
C12	 &  0.124  &  10.01  & 	   4.38  \\
C13	 &  0.213  &  15.05  & 	6.42  \\
c1	 &  0.092  &  20.72  & 		13.56  \\
c2	 &  0.099  & 22.78   & 	9.70  \\
f1	 &  0.511  &   6.76   & 	3.48 \\
f2	 &  0.488  & 	7.21   &   3.88   \\
f3	 &  0.274  & 10.60   & 	1.87  \\
f4	 &  0.509  & 13.80   & 	3.16  \\
\hline
\end{tabular}
\caption{Column density from the ammonia and the H$_2$ maps 
for the G14 sample defined with dendrograms.
 } 
\label{tab:columndensity}
\end{table}

%
 \begin{equation}
	 X_2 = \frac{N(NH_3)}{N(H_2)} = 4.7\times 10^{-8},
 \end{equation}
\begin{figure}[h]
\centering
\includegraphics[width = 12cm]{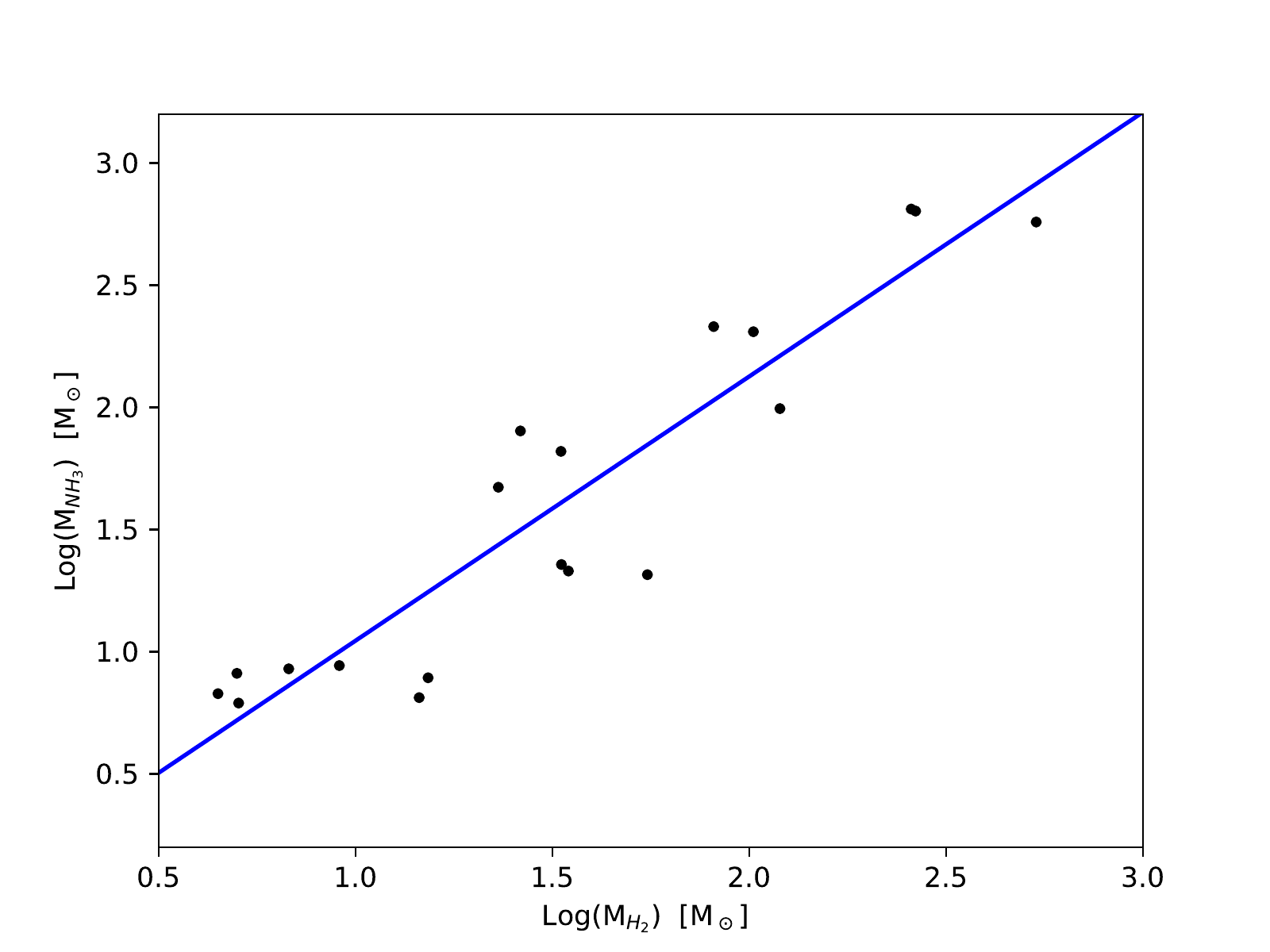}
\caption{Comparison of the objects' mass measured through the NH$_3$ 
data (using $X_1=4.2\times10^{-8}$) and directly from the H$_2$ map. 
The blue line corresponds to the fit, which has a slope of 1.08, showing 
a good correlation for the mass measurements with the different tracers. }
\label{fig:g14masses}
\end{figure}

In order to compute the N(H$_2$) column density from the ammonia data 
we need to assume an abundance, X(NH$_3$). One adopted value in the 
literature, obtained as an average of several samples is 
$X_1= 4.2\times10^{-8}$ \citep[][and references therein]{SanchezM13}. 
However, given the NH$_3$ and H$_2$ column densities, we can compute 
the NH$_3$ abundance for G14, considering the values in Tab. 
\ref{tab:columndensity}. Figure \ref{fig:histoN} shows the comparison for 
the H$_2$ column density obtained directly from the H$_2$ map of 
\citet{Lin17}, the H$_2$ column density obtained using the abundance 
reported in \citet{SanchezM13} and the one obtained using the abundance 
derived in this work. The mass for our G14 sample was computed 
considering $X_1$. Figure \ref{fig:g14masses} shows the comparison 
between the masses derived from $\ammonia$ and the H$_2$ maps. We 
found quite similar and consistent masses using both independent 
methods.

\section{Estimating the impact of neglecting stellar feedback} 
\label{app:feedback}

As mentioned in Sec.\ \ref{sec:caveats}, our simulation does not include any form of stellar feedback. Here we estimate the likely impact of this omission in the realism of our ``numerical cloud". An important consideration in this regard is that the masses of the sink particles allow us to estimate the masses of the typical stars acting on the cloud, and these masses increase over time, representing the growth of the stellar cluster formed by the cloud. Indeed, although our simulation lasts 34 Myr, during an important fraction of that time there is no star formation. The first sink forms only at  $t \sim 16$ Myr, with a mass $\sim 12 \Msun$, and in a region  $\sim 60$ pc from region A. In the region A itself, the first local sink appears at $t \sim 21.6$ Myr, with mass  $\sim 3 \Msun$. For comparison, we study our cloud within region A in the time interval from $t_{\rm i} = 20.3$ Myr to $t_{\rm f} = 22.1$ Myr, so that, in fact, the region contains no stars at the beginning of our study.

We now evaluate the impact of omitting stellar feedback. If feedback were present, it would have to affect the dynamics of the cloud during the $\sim 2$ Myr we are studying. Some of the most important possible sources of feedback are, from weakest to strongest, {\it i)} outflows, {\it ii)} photoionizing radiation, and {\it iii)} supernovae. Concerning outflows, it is generally agreed that outflows from low-mass YSOs only affect their immediate surroundings, up to distances $\lesssim 1$ pc, but hardly affect their entire parent MC \citep[][and references therein]{Bally16}. Moreover, even in the presence of outflows, the accretion flow onto the cores and YSOs is not prevented \citep[e.g.][]{Wang10} and, in fact, the outflows last only as long as the accretion does \citep{Bally16}. On the other hand, although very massive sources are able to provide powerful outflows that extend over tens of parsecs, they still cannot halt most clumps \citep{Bally16}. Thus, we can discard outflows as relevant for our study.  Several numerical studies \citep[e.g.,] [] {Gavagnin17, VS17, Dale17, Grudic19} also support the suggestion that feedback at the cloud scale is dominated by massive stars, but these appear a few Myr after the onset of SF in the cloud. In further support of this conclusion, we note that the simulations presented in \citet{GS20}, which do include stellar feedback and are performed with a grid code, the general evolution of clouds with and without feedback is nearly identical until a very massive ($\sim 20 \Msun$) star forms.

Concerning photoionization and SNe, we can estimate the severity of the omission as follows. Applying a standard IMF to our sink particles, we find that, at $t = t_{\rm f}$, the most massive star that can be hosted by the most massive sink in our cloud would have a mass $\sim 4 \Msun$. If we consider all the sink particles within our region then the most massive star may have $\sim 7 \Msun$. Such a star does not produce significant photoionizing radiation and does not explode as a SN in at least $\sim 30$ Myr. 

A nonlocal SN explosion could still affect our cloud if it occurs sufficiently nearby in space and time. The nearest star-forming region is $\sim 60$ pc away, and it started forming stars $\sim 6$ Myr before $t_{\rm i}$. For a SN in this region to impact our cloud, it should explode during this 6-Myr interval. For this to happen, the star should have a mass of at least 20$\Msun$. Again applying the standard IMF, a sink must have 2000 $\Msun$ to form at least one $20 \Msun$ star. However, at $t=t_{\rm i}$, the neighbouring cluster has a total mass of only $\sim 1550 \Msun$, which is not enough to form a $ 20 \Msun$ star. Conversely, it only reaches a total mass of $\sim 2000$Msun at $t \sim 21.6$ Myr, the moment when the first sink particle appears in the cloud. We stop the analysis just 0.5 Myr after that. Thus, even though the neighboring region eventually reaches a sufficiently large mass to form the required massive star, it does not have time to interact with our region.

Finally, a last possibility exists: that of a passing Type Ia SN, that could explode in the neighborhood of our cloud. The standard estimate is that a given cloud is hit by a SN shock roughly once every 1 Myr \citep{McKee77}. So, at most, during the evolution of our cloud, one SN shock may pass through it. However, in order for it to cause much damage to the cloud, the SN must explode inside it, while external explosions cause very little damage \citep[e.g.][]{Iffrig15}. Therefore, we conclude that the omission of stellar feedback does not pose a significant problem for the evolution of our cloud’s dynamics.



\listofchanges

\end{document}